\documentclass[aps,prx,amsmath,floats,floatfix,twocolumn,
  superscriptaddress,nofootinbib,showpacs]{revtex4-1}

\pdfoutput=1 
\usepackage{diagbox}
\usepackage{multirow}
\usepackage{braket}
\usepackage{amsfonts}
\usepackage{amsmath}
\usepackage{amssymb}
\usepackage{amsthm}
\usepackage{bm}
\usepackage{graphicx}
\usepackage{graphics}
\usepackage{latexsym}
\usepackage{rotating}
\usepackage[dvipsnames]{xcolor}
\usepackage{mathrsfs}
\usepackage{microtype}
\usepackage{verbatim}
\usepackage{url}

\usepackage[caption=false]{subfig}
\usepackage[normalem]{ulem}

\usepackage[breaklinks=true]{hyperref}
\hypersetup{
    colorlinks=true,
    linkcolor=NavyBlue,
    filecolor=Magenta,
    urlcolor=NavyBlue,
    citecolor=NavyBlue
}

\usepackage{yfonts}
\usepackage{float}
\usepackage{xspace} 
\usepackage{mathrsfs}
\usepackage{siunitx}
\usepackage{array}

\newcommand{\mL}{\mathcal{L}}
\newcommand{\baq}{\bar{\alpha}_1}
\newcommand{\baw}{\bar{\alpha}_2}

\newcommand{\fdamp}{f_{\text{damp}}}
\newcommand{\fRD}{f_{\text{RD}}}
\newcommand{\lbdi}[1]{\lambda^i_{#1}}
\newcommand{\hp}{h^\prime}

\newcommand{\la}{\lambda^a}
\newcommand{\lam}{\lambda^a_{\max}}
\newcommand{\rhm}{\rho_{\max}}

\newcommand{\rtrmrp}{\frac{\rho_{\max}}{\rho^\prime}}

\newcommand{\Ione}{R_{\mu\nu\ \ }^{\ \ \alpha\beta}R_{\alpha\beta\ \ }^{\ \ \gamma\sigma}R_{\gamma\sigma\ \ }^{\ \ \mu\nu}}
\newcommand{\Itwo}{R_{\mu\ \nu\ }^{\ \alpha\ \beta} R_{\alpha\ \beta\ }^{\ \gamma\ \sigma} R_{\gamma\ \sigma\ }^{\ \mu\ \nu}}
\newcommand{\ttt}[1]{\text{#1}}
\newcommand{\mc}[1]{\mathcal{#1}}

\begin{document}

\title{Robust and improved constraints on \\ higher-curvature gravitational effective-field-theory with the GW170608 event}

\author{Haoyang Liu}
\affiliation{Illinois Center for Advanced Studies of the Universe \& Department of Physics, University of Illinois at Urbana-Champaign, Urbana, Illinois 61801, USA}

\author{Nicol\'as Yunes}
\affiliation{Illinois Center for Advanced Studies of the Universe \& Department of Physics, University of Illinois at Urbana-Champaign, Urbana, Illinois 61801, USA}

\begin{abstract}
Effective field theory methods allow us to modify general relativity through higher-curvature corrections to the Einstein-Hilbert action, while preserving Lorentz invariance and the number of gravitational degrees of freedom. 
We here construct an approximate inspiral-merger-ringdown waveform model within the cubic, parity-preserving class of effective-field-theory extensions to Einstein's theory for the gravitational waves emitted by quasi-circular binary black holes with aligned/anti-aligned spins.
Using this waveform model, we first explore the detectability of non-Einsteinian effective-field-theory effects through an extended version of effective cycles to illustrate the need to include non-Einsteinian amplitude corrections.
We then use this model to analyze the GW170608 event in a full Bayesian framework, and we place new improved and more robust constraints on the coupling constants of the effective field theory. 
Our Bayesian model selection study disfavors the non-Einsteinian theory with a (log) Bayes factor of $\log \mathcal{B}^{\text{EFT}}_{\text{GR}} = -2.81$. 
Our Bayesian parameter estimation study places the constraints 
$\bar{\alpha}_1=0.87^{+1.95}_{-1.03}$ and $ \bar{\alpha}_2=-0.35^{+4.12}_{-2.92}$ 
at $90\%$ confidence on the coupling parameters of the effective-field theory. 
These constraints are $3.5$ stronger than previous constraints, informative relative to the prior, and independent of the choice of prior on the coupling parameters of the modified theory. 

\end{abstract}

\maketitle

\section{Introduction}

The detection of gravitational waves (GWs) has given us powerful tools to probe gravitational physics in the highly dynamical, strong field regime, such as when black holes (BHs) collide and neutron stars merge~\cite{LIGOScientific:2018mvr, LIGOScientific:2020ibl, KAGRA:2021vkt}. Within the past few years, more than 100 events have been observed by current, ground-based GW detectors~\cite{KAGRA:2021vkt}.
These observations have provided new opportunities to learn about the properties of gravity and compact objects in such extreme conditions. In particular, these events allow us to study the possible existence of nonbaryonic matter in compact objects and to investigate the properties of the gravitational theory in play during compact object coalescence. 

Many tests have shown that Einstein's theory is currently the most successful gravitational description of nature, but some theoretical problems and observational anomalies may suggest otherwise. On the theoretical side, GR's incompatibility with quantum mechanics \cite{Shomer:2007vq} and the singularity problem \cite{PhysRevLett.14.57} remain unsolved problems. On the observational side, the late-time acceleration of the universe~\cite{SupernovaSearchTeam:1998fmf,SupernovaCosmologyProject:1998vns,Planck:2018nkj,WMAP:2012nax}
 and galaxy rotation curves~\cite{1914LowOB...2...66S} continue to be anomalies, unless one invokes an unnaturally small cosmological constant~\cite{Carroll:2000fy} and a large amount of dark matter~\cite{Bertone:2004pz,Bertone:2016nfn}. Since Solar-system observations indicate GR is right in the weak field~\cite{Will:2014kxa}, deviations that attempt to solve the above problems may occur in the strong and highly-dynamical regime, such as in the merger of black holes. Therefore, the observation of GWs emitted in such coalescence events provide a new window for searching for or constraining modifications of Einstein's theory~\cite{Yunes:2013dva}. 

Many theories have been proposed to extend GR. For example, Scalar-Tensor (ST) theory \cite{Damour_1992,PhysRevLett.70.2220} introduces an extra scalar degree of freedom that couples non-minimally to the metric tensor. Other theories assume the existence of higher spacetime dimensions~\cite{Arkani-Hamed:1998jmv,Randall:1999ee,Randall:1999vf} or non-local interactions in the action~\cite{PhysRev.77.219,PhysRev.80.1047}. On the other hand, if we believe GR has to be modified above a certain energy or curvature scale, but we wish to remain agnostic about interactions at higher energies or curvatures, we can then use effective-field theory (EFT) techniques~\cite{Donoghue:1994dn, Donoghue:1995cz} to extend GR. An EFT extension of GR that includes a series of higher-order curvature terms in the action, without the inclusion of additional scalar or vector degrees of freedom, has recently been proposed and analyzed in the context of GWs and BHs~\cite{Endlich:2017tqa,Cardoso:2018ptl,Sennett:2019bpc,AccettulliHuber:2020dal,deRham:2020ejn}. This extension differs from others, such as dynamical Chern-Simons gravity~\cite{ALEXANDER20091}, Einstein-dilaton-Gauss-Bonnet  gravity\cite{Moura_2007,PhysRevD.79.084031}, and Einstein-Scalar-Gauss-Bonnet gravity \cite{Weinberg:2008hq,Kovacs:2020pns}, in that the higher-curvature corrections are not multiplied by a function of a new scalar field. Instead, the higher-curvature corrections of the EFT are simply multiplied by coupling constants that select the type and strength of the higher-curvature interactions considered. 

We here investigate these higher-curvature EFT extensions of GR in light of recent GW observations. In particular, we focus on higher-curvature EFT corrections that preserve Lorentz symmetry and do not possess additional degrees of freedom beyond those contained in GR. Previous work placed constraints on this class of EFT extensions using inspiral-only models~\cite{Sennett:2019bpc,Liu:2023spp} or inspiral-merger-ringdown waveform models, with the EFT corrections included only in the ringdown only~\cite{Silva:2022srr}. One of the main results of this paper is the extension of these studies by developing an approximate inspiral-merger-ringdown (IMR) waveform that is capable of modeling the entire coalescence of BHs in this class of theories. In particular, we introduce EFT modifications to the (Fourier GW amplitude and phase of the) IMRPhenomD waveform model in the inspiral and ringdown stages of coalescence, treating the merger phase as an interpolating region. In order to accomplish this, we first re-compute the EFT modifications to the inspiral waveform to 1 post-Newtonian (PN) order\footnote{The PN approximation is one in which the Einstein equations are solved perturbative in small velocities and weak fields; a term of APN order scales as $v^{2A}$ relative to the controlling factor of the expansion.} beyond the leading PN order EFT modification (which actually enters at 5PN order), correcting a few inaccuracies that have appeared in the literature. Then, we correct the ringdown waveform by adopting fitting functions for the EFT corrections to the dominant $(n,l,m) = (0,2,2)$ QNM frequencies that are valid up to dimensionless remnant BH spins of $\chi < 0.7$. Even though this IMR model is obviously not complete (as it lacks the inclusion of direct EFT modifications to the merger, or a complete description of the 6PN EFT terms), it still stands as the first and most accurate EFT model for the IMR of BHs in this theory constructed to date. 

Since the modifications to the GW amplitude and phase in the inspiral and ringdown all depend on the same coupling constants, comparisons of this IMR model against the data should lead to stronger and more robust constraints on this class of theories. Another main result of this paper is our confirmation of this expectation through the use of the new IMR model to analyze GW signals and place new constraints. We first extend the concept of ``effective cycles'' to include amplitude corrections and use this extended concept to assess the detectability of non-GR EFT effects. We find that GW amplitude corrections in the inspiral are as important (if not more important) than GW phase corrections for the EFT corrections considered here. This is because the non-GR effects enter at relativity high PN order, becoming important in the very last stages of inspiral, where the GW amplitude does not play a subdominant role. We then carry out a synthetic injection and recovery campaign using Bayesian model selection and find that an SNR of above $\sim 18$ is sufficient for the EFT IMR model to be preferred over the GR model.

Finally, we use our new IMR model to analyze real GW data, focusing on the GW170608 event, and using Bayesian parameter estimation. We find that the data prefers the GR model, allowing us to place new stringent constraints on the EFT coupling parameters that are about a factor of 5 more stringent than previous constraints. In particular, we are careful to create priors that ensure our IMR model remains valid in the analysis of GW data (i.e.~that the IMR model remains within the cutoff of the EFT theory), and we demonstrate explicitly that our posteriors on the EFT coupling parameters are \textit{independent} of the choice of prior. This ensures the validity of our Bayesian analysis and the robustness of our new constraints. 

The rest of the paper describes the above results in more detail and is organized as follows. 
In Sec.~\ref{sec.II}, we introduce the cubic EFT theory and its effects on the orbital binding energy and the GW flux. 
In Sec.~\ref{sec.III}, we explain how to construct our new IMR waveform model. We first review the construction of the IMRPhenomD model in GR [Sec.~\ref{sec.III:A}], and then we discuss how to derive the EFT corrections to the GW model and how to incorporate them into the GR model [Sec.~\ref{sec.III:B}]. 
In Sec.~\ref{sec.IV}, we investigate the priors we must impose on the parameters of our new IMR model to ensure the model's validity. 
In Sec.~\ref{sec.V}, we discuss the detectability of EFT effects described by our model through effective cycles and through tests with synthetic injections and recoveries. Finally, in Sec.~\ref{sec.VI}, we present the result of our Bayesian analysis on real GW data for the GW170608 event. We conclude and point to future research in Sec.~\ref{sec.VII}.  Appendix~\ref{sec:appendix} presents the details of the calculation of the orbital energy and the energy flux in the inspiral regime to 1PN order, correcting some mistakes found in the literature. Appendix~\ref{sec:appendix_B} briefly introduces how to estimate the 5PN coefficient through extrapolation techniques.

\section{Cubic Effective Field Theory Extension of GR}
\label{sec.II}

The cubic EFT extension of GR is an effective field theory that attempts to describe gravity below a specific (cut-off) energy scale $\Lambda$. In this work, we consider an EFT that preserves Lorentz invariance and does not introduce extra degrees of freedom. In particular, we consider the action
\begin{equation}\label{Eq. action}
    S = \frac{1}{16\pi G}\int d^4 x \sqrt{-g} \left[R + \frac{\mL_{D4}}{\Lambda^2} + \frac{\mL_{D6}}{\Lambda^4} + ... \right]
\end{equation}
where $\mL_{Dn}$ stands for all the $n$-dimension operators that can be constructed from the Riemann curvature, its contractions, and its derivatives.
The Lagrangian of the dimension-4 term can be written as
\begin{equation}
    \mL_{D4} = a_{GB} R^2_{GB} + a_{R^2}R^2 + a_{2}R^{\mu\nu}R_{\mu\nu}
\end{equation}
where $R^2_{GB} = R^2_{\mu\nu\alpha\beta} - 4 R^2_{\mu\nu} + R^2$ is the Gauss-Bonnet term, which is topological in 4 dimensions. The remaining $R^2$ and $R^{\mu\nu}R_{\mu\nu}$ terms in the action will not affect the equations of motion within an EFT treatment, because they introduce modifications to the Einstein equations that depend on $R$ or $R_{\mu \nu}$, which are zero in GR for black hole binaries~\cite{deRham:2020ejn}. 
Therefore, $\mL_{D4}$ will not contribute and we ignore this term in our work. 

The dimension-6 term in the Lagrangian can be written as
\begin{equation}\label{Eq. L_D6}
    \mL_{D6} = \alpha_1^{\text{EFT}} I_1 + \frac{\alpha_2^{\text{EFT}}}{2} (I_1 - 2 I_2)
\end{equation}
where $I_1 = \Ione $ and $I_2 = \Itwo $. The above interactions are proportional to two EFT coupling constants, $\alpha_1^{\text{EFT}}$ and $\alpha_2^{\text{EFT}}$, which we non-dimensionalize via 
\begin{equation}\label{eq.ba12}
    \bar{\alpha}_{1,2} = \frac{c^8}{(GM\Lambda)^4}\alpha_{1,2}^{\text{EFT}}\,.
\end{equation}
The above quantity is dimensionless because $\alpha_{1,2}^{\text{EFT}}$ is assumed to be dimensionless (and of order unity), while $M$ is the total mass and has units of length (in geometric units) and $\Lambda$ has units of energy (in natural units) or inverse length (in geometric units). 
Unlike $I_1$, the combination $I_1 - 2 I_2$ does not contribute to isolated bodies in vacuum~\cite{Brandhuber:2019qpg}; however, it affects the binding energy~\cite{Brandhuber:2019qpg,Emond:2019crr} and the gravitational radiation~\cite{AccettulliHuber:2020dal} emitted by two objects in a binary system, even if the objects are BHs. Thus, the $I_1 - 2 I_2$ term only appears in the interactions of two-body system and can be neglected for isolated BHs.

The truncation of the EFT at order 6 is consistent, provided the terms dropped can be ignored. This is expected to be the case provided the cubic EFT terms dominate the higher-order EFT terms neglected. Thus, we need $\left[ R \right]/\Lambda^2 \ll 1$ where $\left[ R \right]$ is the typical size of the Riemann curvature. For two objects in a binary system, we can approximate this quantity via $\left[ R \right] = GM/c^2 r_{12}^3$, where $r_{12}$ is the separation between two objects. Thus, we arrive at
\begin{equation}
    f \ll c\Lambda/\pi\,,
\end{equation}
which is the EFT cut-off in the frequency band. This cutoff is slightly different from that which appeared previously in the literature~\cite{Sennett:2019bpc}, because we model the characteristic size of the Riemann curvature with that of a binary system, instead of that of an isolated black hole. In this work, since we focus on the performance of our consistent EFT waveform model and the cutoff $\Lambda$ has been absorbed in $\bar{\alpha}_i$, we will assume that the whole GW signal in the detector band is well below the cutoff, i.e.~we will assume that $\Lambda \gg \pi f_{\rm max} \approx 3 \times 10^3 \; {\rm{Hz}} \approx 2 \times 10^{-12}$ eV, where $f_{\rm max} \sim 10^3$ Hz is the approximate maximum frequency that the advanced LIGO detectors were sensitive to during the third observing run.

\section{Waveform Construction}
\label{sec.III}

To construct an approximate IMR waveform, we calculate the EFT corrections  to the GW amplitude and phase in the stationary phase approximation (SPA). We then add these corrections to the GR IMR waveform, and in this paper, we model the GR sector with the IMRPhenomD model. In this section, we first introduce the IMRPhenomD model and, specifically, we explain how different stages of the coalescence are stitched together. Then, we derive the EFT corrections to the GW phase and amplitude in the inspiral and ringdown stages.

\subsection{IMRPhenomD waveform}
\label{sec.III:A}

IMRPhenomD\cite{Husa:2015iqa, Khan:2015jqa} is a phenomenological, spin-aligned, binary BH (BBH) waveform model, which only accounts for the dominant $(2,2)$ GW mode. This model was developed within GR, and we here review some of its more salient features. To avoid notational clutter, we refrain from labeling every quantity in this section with a ``GR'' subscript, with the understanding that, as mentioned earlier, the IMRPhenomD model has been developed within Einstein's theory. 

The IMRPhenomD GW model is defined in the Fourier domain via
\begin{equation}
\label{eq:IMRPhenomD-model}
    \tilde{h}(f) = A(f)e^{i\phi(f)}\,,
\end{equation}
and it models the GW phase and amplitude separately. These quantities are prescribed as piecewise functions that represent the inspiral stage, the intermediate stage, and a merger-ringdown stage, according to 
\begin{equation}\label{Eq. IMRPhD_phase}
    \phi (f) = \begin{cases}
        \phi_{\text{Ins}}(f), & f \leq f^{\text{phase}}_1 \\
        \phi_{\text{Int}}(f), & f^{\text{phase}}_1 < f < f^{\text{phase}}_2 \\
        \phi_{\text{MR}}(f), & f \geq f^{\text{phase}}_2 
    \end{cases}
\end{equation}
and
\begin{equation}\label{Eq. IMRPhD_phase}
    A(f) = \begin{cases}
        A_{\text{Ins}}(f), & f \leq f^{\text{amp}}_1 \\
        A_{\text{Int}}(f), & f^{\text{amp}}_1 < f < f^{\text{amp}}_2 \\
        A_{\text{MR}}(f), & f \geq f^{\text{amp}}_2 
    \end{cases}
\end{equation}
The transition frequencies between the inspiral and intermediate stage are set to $f^{\text{phase}}_1 = 0.018/M$ and $f^{\text{amp}}_1 = 0.014/M$, where $M = m_1 + m_2$ is the total mass. The transition times between the intermediate and the merger-rindown stage are set to $f^{\text{phase}}_2 = 0.5 f_{\text{RD}}$ and $f^{\text{amp}}_2 = f_{\text{peak}}$, where
\begin{equation}\label{Eq. f_peak}
    f_{\text{peak}} = \left| f_{\text{RD}} + \frac{f_{\text{damp}} \gamma_3 (\sqrt{1-\gamma_2^2}-1) }{\gamma_2} \right|
\end{equation}
$\gamma_i$ are phenomenological parameters, and $f_{\text{RD}}$ and $f_{\text{damp}}$ are the ringing and damping frequencies of the remnant BH.

In the inspiral stage, the GW phase is modeled through a 3PN TaylorF2 approximant $ \phi_{\text{TF2}}$ (see e.g.~\cite{PhysRevD.80.084043,Bohe:2013cla,Poisson:1997ha}), 
enhanced with four fitting coefficients $\sigma_i$, namely
\begin{align}\label{Eq. IMPhD_Ins_phase}
    \phi_{\text{Ins}}(f) &= \phi_{\text{TF2}}(f) 
    \nonumber \\
    &+ \frac{1}{\eta} \left(\sigma_0 + \sigma_1 f + \frac{3}{4}\sigma_2 f^{4/3} + \frac{3}{5} \sigma_3 f^{5/3} + \frac{1}{2}\sigma_4 f^2\right)\,,
\end{align}
where $\eta = m_1 m_2/(m_1 + m_2)^2$ is the symmetric mass ratio for a binary system with masses $m_1$ and $m_2$.
The GW amplitude follows a similar logic, and it is thus written in terms of a re-expanded 3PN SPA amplitude $A_{\text{PN}}$ (see e.g.~\cite{Khan:2015jqa}) that is enhanced with four fitting coefficients $\rho_i$, namely
\begin{equation}\label{Eq. IMRPhD_Ins_amp}
    A_{\text{Ins}}(f) = A_{\text{PN}}(f) + A_0 \sum^3_{i=1} \rho_i f^{(6+i)/3}\,,
\end{equation}
with $A_0 = \sqrt{{2 \eta}/({3\pi^{1/3}})}f^{-7/6}$. The fitting coefficients $\sigma_i$ and $\rho_i$ are calibrated to hybrid data from the SEOBNRv2 model and NR simulations~\cite{Husa:2015iqa, Khan:2015jqa}, as we will discuss later. 

In the intermediate stage, the GW phase is given by
\begin{equation}
    \phi_{\text{Int}}(f) = \frac{1}{\eta}\left(\beta_0 + \beta_1 f + \beta_2 \log (f) - \frac{\beta_3}{3}f^{-3}\right)\,,
\end{equation}
while the GW amplitude is 
\begin{equation}
    A_{\text{Int}} = A_0 (\delta_0 + \delta_1 f + \delta_2 f^2 + \delta_3 f^3 + \delta_4 f^4)\,,
\end{equation}
where $\beta_i$ and $\delta_i$ are phenomenological fitting coefficients. 

In the merger-ringdown phase, the GW phase is modeled through 
\begin{equation}\label{Eq. IMRPhD_RD_phase}
\begin{aligned}
    \phi_{\text{MR}} & = \frac{1}{\eta}\left[\alpha_0 + \alpha_1 f - \alpha_2 f^{-1} +\frac{4}{3} \alpha_3 f^{3/4} \right. \\
    & \left. + \alpha_4 \tan^{-1} (\frac{f-\alpha_5 f_{\text{RD}}}{f_{\text{damp}}})\right]
\end{aligned}
\end{equation}
where recall that $f_{\text{RD}}$ and $f_{\text{damp}}$ are the (dominant) ringing and damping frequencies of the remnant BH.
The GW amplitude, on the other hand, is different from that in the inspiral and intermediate stages because it is derived from the Fourier transform of an exponentially-damped sinusoidal function, which yields a Lorentzian. 
Therefore, the GW amplitude is modeled through
\begin{equation}\label{Eq. IMRPhD_RD_amp}
    \frac{A_{\text{MR}}} {A_0} = \gamma_1 \frac{\gamma_3 f_{\text{damp}}}{(f-f_{\text{RD}})^2 + (\gamma_3 f_{\text{damp}})^2} e^{-\frac{\gamma_2 (f-f_{\text{RD}})}{\gamma_3 f{\text{damp}}}}
\end{equation}
where $\alpha_i$ and $\gamma_i$ are phenomenological fitting parameters. Note that, in principle, when the GWs emitted during ringdown are modeled as quasi-normal oscillations, one must sum over all $(n,l,m)$ modes (where $n$ is the overtone number, $l$ is the azimuthal number, and $m$ is the magnetic number of the mode); in practice, however, we here consider only the dominant mode for comparable mass binaries, which have $(n,l,m) = (0,2,2)$ numbers.

The IMRPhenomD model depends on certain phenomenological parameters. In the inspiral, these are $(\sigma_i, \rho_i)$, in the intermediate region $(\beta_i, \delta_i)$, and in the merger-ringdown $(\alpha_i,\gamma_i)$. Some of these parameters $(\alpha_{0/1}, \beta_{0/1}, \gamma_{0/1}, \delta_{0/1})$ can be determined by requiring the GW phase and amplitude be $C(1)$ across the interfaces. All other parameters were fitted to the ansatz
\begin{equation}
    \begin{aligned}
        \Lambda^i = & \lbdi{00} + \lbdi{10}\eta \\
                     & (\chi_{\text{PN}} - 1)(\lbdi{01} + \lbdi{11}\eta + \lbdi{21}\eta^2) \\
                     & + (\chi_{\text{PN}} - 1)^2 (\lbdi{02} + \lbdi{12}\eta + \lbdi{22}\eta^2) \\
                     & + (\chi_{\text{PN}} - 1)^3 (\lbdi{03} + \lbdi{13}\eta + \lbdi{23}\eta^2) \\
    \end{aligned}   
\end{equation}
using numerical relativity simulations. In this fitting function, $\Lambda^i$ stands for the phenomenological parameters, $\chi_{\text{PN}}$ is the (unscaled) reduced-spin parameter, and $\lambda^i$ are fitting coefficients, given in \cite{Khan:2015jqa}.

\subsection{EFT correction to GW}
\label{sec.III:B}

Let us now discuss how EFT corrections modify the IMRPhenomD model in each of the stages of coalescence.

\subsubsection{Inspiral}
\label{sec.III:B.1}

The EFT modifications to the Lagrangian result in a modification to the orbital energy of the system (due to corrections in the Hamiltonian) and to an enhanced amount of GW radiation (and thus, an enhanced GW flux) in the inspiral stage relative to the GR expectation. In the inspiral, the orbital binding energy can be written as $E(v) = E_{\text{GR}} + E_{\text{EFT}}$, where the GR term is (see e.g.~\cite{maggiore2007gravitational})
\begin{align}
 E_{\text{GR}} &= -\frac{1}{2}\eta v_f^2 + \frac{\eta}{24}(9+\nu)v_f^4 + {\cal{O}}(1/c^4)\,,
\end{align}
and the EFT correction is
\begin{equation}\label{Eq. EF_EFT}
    \begin{aligned}
        E_{\text{EFT}} & = \frac{27 \eta \alpha_2^{\text{EFT}}}{(GM\Lambda)^4} v_f^{12} \\
        &- \frac{33\eta \left[ 4\alpha_1^{\text{EFT}} + (2\eta - 11)\alpha_2^{\text{EFT}}\right] }{2(GM\Lambda)^4} v_f^{14}
    \end{aligned}
\end{equation}
where $v_f=(\pi GMf/c^3)^{1/3}$ and $f$ is the GW frequency. We computed this 1PN correction to the binding energy by correcting Kepler's third law with 1PN corrections, and using this in the 1PN-corrected gravitational potential, as we explain in Appendix~\ref{sec:appendix}. 

Similarly, the GW energy flux can be written as ${\cal{F}}(v_f) = {\cal{F}}_{\text{GR}} + {\cal{F}}_{\text{EFT}}$, where the GR term is (see e.g.~\cite{maggiore2007gravitational})
\begin{equation}
 {\cal{F}}_{\text{GR}} = -\frac{32}{5G}\eta^2 v_f^{10} \left[   1 + \left(-\frac{1247}{336} - \frac{35}{12}\eta\right)v_f^2  \right] + {\cal{O}}(1/c^{14})\,,
\end{equation}
and the EFT correction is 
\begin{equation}
    \begin{aligned}
        \mathcal{F}_{\text{EFT}} = & \frac{4608\eta^2}{5G}\frac{\alpha_2^{\text{EFT}}}{(GM\Lambda)^4}v_f^{20} \\
        & + \frac{768\eta^2 }{5G}\frac{7(1+\nu)\alpha_2^{\text{EFT}} -32\alpha_1^{\text{EFT}}}{2(GM\Lambda)^4}v_f^{22}
    \end{aligned}
\end{equation}
We here computed the EFT correction by using the 1PN-corrected and modified version of Kepler's third law in the 1PN flux equation in the GR sector and the Newtonian flux equation in the EFT sector, as shown in Appendix~\ref{sec:appendix}. Notice that this EFT expression for the flux corrects that of \cite{AccettulliHuber:2020dal} at 1PN order, as also explained in Appendix~\ref{sec:appendix}. Note also that this flux expression is not complete, because the 1PN correction to the flux equation in the EFT sector is not known (and would require new scattering amplitude calculations).  

With this in hand, we can now compute the EFT corrections to the GW amplitude and phase in the SPA.  
The SPA assumes that the Fourier transform of the GW time-domain model is dominated by contributions close to a stationary point $t = t_f$, where the integral has an approximate saddle point. Under this approximation, the Fourier-domain GW model can be written as (see e.g.~\cite{Damour:2000gg})
\begin{equation}\label{Eq. waveform_SPA}
    \tilde{h}^{\rm SPA}(f) = \frac{A(t_f)}{\sqrt{\dot{f}(t_f)}} e^{i\psi_f - \frac{\pi}{4}}\,.
\end{equation}
where the phase $\psi_f$ evaluated at the stationary point is
\begin{equation}\label{Eq. phase_SPA}
    \psi_f = 2\pi f t_{\text{ref}} - \phi_{\text{ref}} + \int^{v_{\text{ref}}}_{v_f} (v_f^3 - v^3)\frac{E^\prime (v)}{\mathcal{F}(v)} dv
\end{equation}
with $E^\prime = dE/df$, and $(t_{\rm ref},\phi_{\rm ref})$ a time and phase reference offset. 

We can now use the previously-presented EFT corrections to the binding energy and GW energy flux to calculate the GW Fourier phase. Inserting the above expressions for the binding energy and GW energy flux into Eq.~\eqref{Eq. phase_SPA}, we obtain 
\begin{align}
\label{eq:EFT-ins-phase}
\phi_{\rm Ins-EFT} = \psi_{\text{GR}} + \psi_{\text{EFT}}\,,   
\end{align}
where $\psi_{\text{GR}}$ is the GW Fourier phase in GR (which can be found e.g.~in~\cite{Husa:2015iqa, Khan:2015jqa}), while the EFT correction is 
\begin{equation}\label{Eq. phase_EFT}
    \begin{aligned}
        \psi_{\text{EFT}} & = a_{\rm 5PN}^{\rm EFT} v_f^{5} +a_{\rm 6PN}^{\rm EFT} v_f^{7} \,,
    \end{aligned}
\end{equation}
where
\begin{align}\label{Eq. aPN_EFT}
     a_{\rm 5PN}^{\rm EFT}  &= -\frac{351}{8\eta} \baw 
     \\
     \label{Eq. aPN_EFT_2}
     a_{\rm 6PN}^{\rm EFT} &= \frac{549360 \baq - 45(43683+12908 \eta)\baw}{12544\eta}\,.
\end{align}
Note that we need to include the 1PN term in $E_{\text{GR}}$ and $\mc{F}_{\text{GR}}$ to properly calculate the EFT correction to the GW Fourier phase to 1PN order. These 1PN terms in the GR sector will contribute to the 6PN EFT term through the product of 5PN EFT terms and 1PN GR terms when expanding the integral. These terms were not accounted for in previous work~\cite{AccettulliHuber:2020dal,Liu:2023spp}, so we correct this here.  

Following a similar procedure, we can compute the EFT corrections to the GW Fourier amplitude, which will contain two EFT contributions: one from the time-domain amplitudes evaluated at the stationary point $A(t_f)$, and another one from square root of the reciprocal of the time derivative of the GW frequency. 
For the first term, we use the fact that the time-domain GW amplitude is 
\begin{align}
    A(t) =  \frac{2 G M \eta}{D_L} (r_{12} \Omega)^2\,,
\end{align}
where $D_L$ is the luminosity distance, recall that $r_{12}$ is orbital separation and $\Omega$ is the orbital angular velocity. Using the EFT-modified version of Kepler's third law (see Eqs.~\ref{Eq. Omega2_n} and \ref{Eq. gamma_x} in Appendix~\ref{sec:appendix}), we can convert the above expression into one that depends only on the GW frequency, using that for the dominant GW models $\Omega = \pi f.$ 
For the second term, we use the chain rule $\dot{f}  = {df}/{dt} = ({1}/{\pi}) ({d \Omega}/{dt})$ to rewrite ${1}/{\sqrt{\dot{f}(t_f)}} \propto ({{\pi}/{\dot{\Omega}}})^{1/2}$.
Putting these two corrections together, we then have that the GW Fourier amplitude is
\begin{align}
\label{eq:EFT-ins-amp}
    A_{\rm Ins-EFT} = A_{\ttt{GR}}+A_{\ttt{EFT}}\,,
\end{align}
where $A_{\ttt{GR}}$ is the GW Fourier amplitude in GR (see e.g.~\cite{Husa:2015iqa, Khan:2015jqa}),
while the EFT correction is
\begin{equation}\label{Eq. amp_EFT}
\begin{aligned}
    A_{\text{EFT}} & = A_\text{Newt} \times \\
    &\left[-198 \bar{\alpha}_2 v_f^{10} + 3 \frac{22624 \bar{\alpha}_1 - (53149+16660\eta)\bar{\alpha}_2}{112} v_f^{12} \right]\,.
\end{aligned}
\end{equation}
where 
\begin{equation}
    A_\text{Newt} = \sqrt{\frac{5\pi}{24}} \frac{(GM)^2\sqrt{\eta}}{2 D_L c^5}v_f^{-7/2}
\end{equation}

With the above modifications, we can then define an EFT-corrected IMRPhenomD model in the inspiral stage. We simply use Eq.~\eqref{eq:IMRPhenomD-model}, but we replace $\phi_{\rm Ins}$ with Eq.~\eqref{eq:EFT-ins-phase} and $A_{\rm Ins}$ with Eq.~\eqref{eq:EFT-ins-amp}. 

\subsubsection{Ringdown}
\label{sec.III:B.2}

The EFT corrections to the Einstein-Hilbert action also introduce modifications to the quasi-normal mode (QNM) frequencies of the ringdown stage, relative to the GR expectation. These modifications enter the waveform model mainly through the QNM frequencies in the ringdown stage of the IMRPhenomD model, namely through $\fRD$ and $\fdamp$ in Eq.~(\ref{Eq. IMRPhD_RD_phase}) and Eq.~(\ref{Eq. IMRPhD_RD_amp}). 

Previous work \cite{Cano:2021myl} calculated the EFT corrections to the QNM frequencies for remnant BHs in the slow-rotating approximation. In this calculation, one starts from the metric perturbation and one derives the corrections to QNM frequencies by solving a modified Regge-Wheeler equation to linear order in spin. Reference~\cite{Cano:2021myl} worked explicitly with the cubic EFT action,
\begin{align}
\label{Eq. action_RD}
     S = \frac{1}{16\pi G}\int d^4 x \sqrt{-g} 
     & \left[ R +  \ell^4 
     \left(\lambda_{\rm even} \Ione 
     \right. \right.
    \nonumber \\ 
    & \left. \left. 
    + \lambda_{\rm odd} R_{\mu\nu\ \ }^{\ \ \alpha\beta}R_{\alpha\beta\ \ }^{\ \ \gamma\sigma}\tilde{R}_{\gamma\sigma\ \ }^{\ \ \mu\nu} \right)\right]\,.
\end{align}
The term $R_{\mu\nu\ \ }^{\ \ \alpha\beta}R_{\alpha\beta\ \ }^{\ \ \gamma\sigma}\tilde{R}_{\gamma\sigma\ \ }^{\ \ \mu\nu}$ is parity-breaking and so we will neglect it in this paper by setting $\lambda_{\rm odd} = 0$. The term proportional to $\lambda_{\rm even}$ is the same as that which appears in Eq.~\eqref{Eq. action}, with the identification $\baq=\lambda_{\rm even} \ell^4 / M^4$. With this in mind, the QNM frequencies were calculated to be 
\begin{equation}
\label{eq:QNM-EFT}
    \omega = \omega^{\text{GR}} + \delta \omega\,,
\end{equation}
where $\omega^{\text{GR}}$ is the $(n,l,m) = (0,2,2)$ QNM (complex) frequency in GR (see e.g.~\cite{Berti:2009kk}), while the EFT corrections to this mode are
\begin{equation}\label{Eq. QNM_EFT_a1}
    \begin{aligned}
    \delta \omega^{\pm} = & \frac{c^3 \bar{\alpha}_1}{GM}  \left[0.0533+0.0255i +m\chi \left(0.0812-0.0489i\right)\right]  \\
    & \pm \frac{c^3 \bar{\alpha}_1}{GM} \left[0.0152-0.0556i-m\chi \left(0.155+0.0249i\right) \right]^{1/2}\,. 
    \end{aligned}
\end{equation}
We clearly see that $\delta \omega$ is linear in $\bar{\alpha}_1$, as expected. 

\begin{figure}[t!]
    \centering
    \includegraphics[width=\columnwidth]{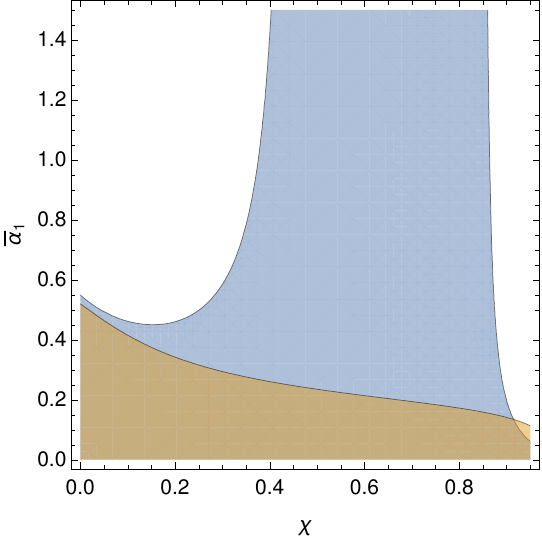}
    \caption{The position in the $\chi-\baq$ plane where the EFT correction flips the sign of the imaginary part of the QNM frequency. The orange and blue regions show where the linear-in-spin approximation of Eq.~(\ref{Eq. QNM_EFT_a1}) ad the higher-in-spin calculation represented by Eq.~(\ref{Eq. QNM_EFT_a12}) are valid, respectively.}
    \label{fig.QNM_chi_ba1}
\end{figure}

\begin{figure}[t!]
    \centering
    \includegraphics[width=\columnwidth]{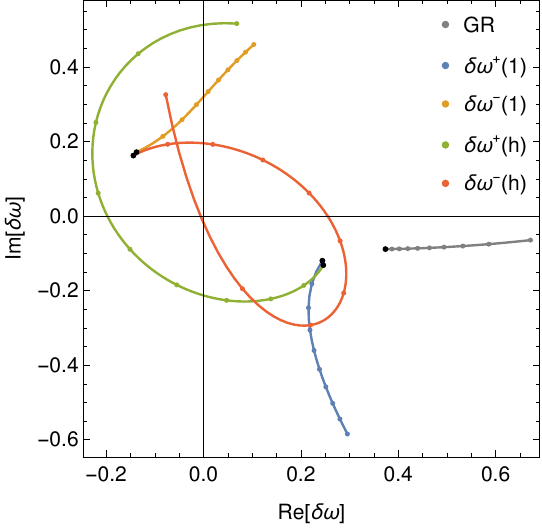}
    \caption{EFT corrections to the QNM frequencies of the $(0,2,2)$ mode with $\baq=1$ (for visual reasons), $\delta \omega^{\pm}$ calculated at linear order in spin ($\delta \omega (1)$) and in the modified Teukolsky approach ($\delta \omega (h)$). The dots indicate an increasing sequence of background spins, starting at $\chi = 0$ (black dot) and increasing up to $\chi = 0.7$ in steps of $\delta \chi = 0.1$. Observe that $\delta \omega^{\pm}(1) \approx \delta \omega^{\pm}(h)$ when $\chi \ll 1$, but they separate when $\chi > 0.1$. For comparison, we also show the $(n,l,m)=(0,2,2)$ QNM frequencies in GR, computed in the Teukolsky approach (gray line). The distance between the colored and the gray lines is proportional to $\bar{\alpha}_1$, and thus, they can be made perturbative by choosing smaller couplings.  }
    \label{fig.QNM_EFT}
\end{figure}

Observe that there are two corrections to the dominant complex frequency: $\delta \omega^+$ and $\delta \omega^-$. 
In GR, quasinormal modes are isospectral, so axial and polar perturbations oscillate with the same frequencies. In modified gravity, however, this need not be the case, and in fact, in general, isospectrality tends to be broken~\cite{Li:2023ulk}. This is indeed what happens in the modified theory we consider here, as shown in~\cite{Cano:2023jbk}. The odd and parity perturbations will both contribute to the plus and cross GW polarizations, and thus, to the response function that an instrument measures. In practice, however, the $\delta \omega^+$ mode damp faster than the $\delta \omega^-$ ones, so, following~\cite{Silva:2022srr}, we keep the latter only. 

However, the above expression in Eq.~\eqref{Eq. QNM_EFT_a1} reveals a problem. When $\bar{\alpha}_1$ is large, $\delta\omega^{(-)}$ can become large enough to overwhelm $\omega^{\rm GR}$ and, in particular, lead to exponentially growing QNMs. Such a linear instability is an obvious sign of the failure of the small-spin expansion, which was used to derive Eq.~\eqref{Eq. QNM_EFT_a1}, and it also indicates that we have exceeded the EFT cut-off. Figure~\ref{fig.QNM_chi_ba1} shows where this happens for the $(n,l, m) = (0,2,2)$ mode. The orange region is where the correction to Eq.~(\ref{Eq. QNM_EFT_a1}) is valid. Observe how the region outside the orange-shaded area is large and excludes a large region of parameter space where the model is not valid.  

A linear-in-spin calculation of the QNM frequencies in modified gravity is not sufficient for GW data analysis, since the remnants of BH collisions are typically not spinning slowly. Recently, a new modified Teukolsky formalism was developed~\cite{Li:2022pcy}, and then used in~\cite{Cano:2023jbk} to derive the QNM frequencies in this EFT theory for BHs with arbitrary spins, which were then used and extended to overtones in \cite{Silva:2024ffz}. The results can be fitted to a polynomial of the form
\begin{equation}\label{Eq. QNM_EFT_a12}
    \delta \omega_{\text{fit}}^{(\pm)} = \baq \sum_{n=0}^N c_n^{(\pm)} \chi^n\,,
\end{equation}
where $N=12$. This polynomial is expected to provide a high-accuracy description of the EFT corrections to the QNM frequencies for BHs with spins up to $\chi \sim 0.7$.

Figure~\ref{fig.QNM_EFT} presents the trajectory of the GR and the EFT correction to the QNM frequencies for the $(n,l,m) = (0,2,2)$ mode, computed both with the linear-in-spin approximation and with the modified Teukoslky approach. Observe that, at small spins, the two calculations are consistent with each other\footnote{The careful observer will notice that $\delta \omega^{\pm}(1)$ is not exactly equal to $\delta \omega^{\pm}(h)$ when $\chi = 0$. This is because both of these calculations have numerical error, and we are plotting fitting functions. This small error will have a negligible effect in our parameter estimation studies of the next section.},
but the trajectories quickly separate, once $\chi \gtrsim 0.1$. Observe also that $\Im(\delta \omega^+) < \Im(\delta \omega^-)$, and thus, the minus branch is the longest-living mode, as also mentioned earlier. 
Henceforth, we will use $\delta \omega^-(h)$ to represent the complex frequencies of the dominant $(n,l,m) = (0,2,2)$ quasinormal mode for all spins. This will mitigate the problems introduced by the linear-order-in-spin approximation, as we can see in Fig.~\ref{fig.QNM_chi_ba1}. The blue region shows where the imaginary part of the correction of Eq. (\ref{Eq. QNM_EFT_a12}) remains smaller than the imaginary part of the GR frequency, therefore remaining within the cutoff of the EFT. Observe how this blue region is much larger than the orange one, except as $\chi \to 1$, where the linear-in-spin approximation is most definitely invalid.  

With the above in mind, we can now define an EFT-corrected IMRPhenomD model in the merger-ringdown stage. We simply use Eq.~\eqref{eq:IMRPhenomD-model}, but we replace $\fRD$ and $\fdamp$ with Eq.~\eqref{eq:QNM-EFT}, using $\delta \omega = \delta \omega^- = \delta \omega_{\rm fit}^-$ with Eq.~\eqref{Eq. QNM_EFT_a12}, with $f_{\rm RD} = \Re(\omega_{\rm GR} + \delta \omega^-_{\rm fit})/2\pi$ and $f_{\rm damp} = \Im(\omega_{\rm GR} + \delta \omega^-_{\rm fit})/2\pi$. 

\section{Model requirements for the prior range}
\label{sec.IV}

The theory we wish to test against GW data is, by definition, an EFT, and thus, as such, it possesses a finite regime of validity. In colloquial terms, one must require that $\baq$ and $\baw$ be ``small enough'' to ensure that the physical scenario one is considering remains within the cutoff scale of the EFT. Moreover, one must ensure that the model that one has developed, which relies on certain approximations (like the PN scheme and BH perturbation theory), remains valid for the range of $\baq$ and $\baw$ considered. In this section, we study and impose these conditions by developing priors on $\baq$ and $\baw$ that we can later implement in a Bayesian analysis of real data. 

The first prior we discuss will ensure the PN description of the inspiral stage of coalescence remains valid.  
The PN description of the modified inspiral remains valid provided the PN series is ``sufficiently convergent,'' i.e.~provided that the EFT corrections are not too large that they shrink the regime of validity of the PN approximation in GR considerably. We follow here the careful investigations of~\cite{Perkins:2022fhr} and require that 
\begin{equation}\label{Eq. PN_prior_1}
    |a^{\text{EFT}}_{\text{5PN}}| > |a^{\text{EFT}}_{\text{6PN}} v^2_{\text{eval}}|
\end{equation}
where recall that $a^{\text{EFT}}_{\text{5PN}}$ and $a^{\text{EFT}}_{\text{6PN}}$ are the coefficients of $v_f^{10}$ and $v_f^{12}$ terms in Eq.~(\ref{Eq. phase_EFT}) and $v_{\text{eval}}$ is the value where we impose Eq.~(\ref{Eq. PN_prior_1}). Following \cite{Perkins:2022fhr}, we take $v_{\text{eval}} = 0.1$ to generate a conservative prior. 

Another prior we impose will ensure that EFT corrections remain small deformations away from GR effects. This prior can be imposed by requiring that the PN coefficient of the EFT correction to the GW phase be smaller than the GR coefficients that enter at the same PN order. Since EFT corrections appear at 5PN and 6PN orders, we then have that
\begin{align}
    \label{Eq. PN_prior_2}|a^{\text{EFT}}_{\text{5PN}}| & < |a^{\text{GR}}_{\text{5PN}}|\,, \\
    \label{Eq. PN_prior_3}|a^{\text{EFT}}_{\text{6PN}}| & < |a^{\text{GR}}_{\text{6PN}}|\,. 
\end{align}
Unfortunately, the 5PN and 6PN coefficients have not yet been calculated in GR. We then model these coefficients by extrapolating the PN sequence in GR, from 0PN to $4.5$ PN order and beyond~\cite{Blanchet:2023bwj}, as shown in Appendix~\ref{sec:appendix_B}. Doing so, 
we find that, for equal mass binaries, 
$a^{\text{GR}}_{\text{5PN}} \sim 1784$ and $a^{\text{GR}}_{\text{6PN}} \sim 8528$,
and we use this in Eq.~(\ref{Eq. PN_prior_2})-(\ref{Eq. PN_prior_3}) to set this prior. 

From Eq.~(\ref{Eq. aPN_EFT}) and Eq.~(\ref{Eq. aPN_EFT_2}) the coefficient $a^{\text{EFT}}_{\text{5PN}}$ and $a^{\text{EFT}}_{\text{6PN}}$ can be written as a function of $\baq$ and $\baw$: $a^{\text{EFT}}_{\text{5PN}} = f_\ttt{5PN}(\baw)$ and $a^{\text{EFT}}_{\text{6PN}} = f_\ttt{6PN}(\baq,\baw)$, and thus, the priors in Eqs.~(\ref{Eq. PN_prior_1}) and (\ref{Eq. PN_prior_2}) can be converted to prior on $\baq$ and $\baw$. More concretely, with $\eta=0.25$, we have that
\begin{align}
    \label{Eq. PN_prior_1_ba}|f_\text{5PN}(\baw)| & > |f_\text{6PN}(\baq,\baw) v^2_{\text{eval}}| \\
    \label{Eq. PN_prior_2_ba} |f_\text{5PN}(\baw)| & < |a^{\text{GR}}_{\text{5PN}}|  \\
    \label{Eq. PN_prior_3_ba} |f_\text{6PN}(\baq,\baw)| & < |a^{\text{GR}}_{\text{6PN}}| 
\end{align}
The prior in Eq.~\eqref{Eq. PN_prior_2_ba} is one-dimensional, but that in Eqs.~\eqref{Eq. PN_prior_1_ba} and~\eqref{Eq. PN_prior_3_ba} is two-dimensional. Therefore, choosing flat priors in $(\baq,\baw)$ is not the same as choosing flat priors in $(a_\ttt{5PN}, a_\ttt{6PN})$, and some choices of the latter will violate the EFT-derived priors above. 
 
Besides these priors that arise from considerations related to the inspiral, we must also impose additional conditions related to the ringdown stage and how this fits within the IMR model. 
First, the IMRPhenomD model must have reasonable boundaries that separate its different frequency regions; we can ensure this is the case, by requiring that $f^{\text{phase}}_2 > f^{\text{phase}}_1$ and $f^{\text{amp}}_2 > f^{\text{amp}}_1 $. Second, the merger-ringdown should occur within the coalescence; we can ensure this is the case by requiring that $f_{\text{RD}} < f_{\max}$, where $f_{\max}$ is the maximum frequency of the waveform model. To ensure the modified waveform is smooth, all of these requirements result in the following prior on $\baq$ and $\baw$: 
\begin{align}
    0.2c^3/(GM) > f_{\text{RD}}(\baq, \baw) > 0.036c^3/(GM) \label{Eq. prior1}\\
    f_{\text{peak}}(\baq, \baw) > 0.015c^3/(GM)\,. \label{Eq. prior2}
\end{align}

Finally, let us discuss the issue mentioned earlier related to the fact that for certain values of $\bar{\alpha}_1$ and the remnant BH spin, the EFT-corrected QNMs may represent exponentially growing GWs. This occurs  when the imaginary part of the EFT correction to the dominant QNM frequency becomes larger than the GR imaginary part. Such a behavior is, of course, unphysical, and thus, we restrict it by imposing a prior on $\bar{\alpha}_1$ and the remnant BH spin $\chi$, namely
\begin{align}\label{Eq. prior_Im}
    \Im\left[\omega^{\rm GR}\left(\chi\right)\right] < -\Im\left[ \omega_{\rm fit}^{(-)}\left(\bar{\alpha}_1,\chi \right) \right]\,.
\end{align}
Further, the ringdown waveform should not exceed the length of the post-merger data, which is typically $2s$ (at most); we can ensure this is not the case by requiring that 
\begin{equation}\label{Eq. prior_fdamp}
    \fdamp > 0.5 \text{Hz}
\end{equation}
The blue region in Fig.~\ref{fig.QNM_chi_ba1} presents the region in the $\bar{\alpha}_1$ and remnant $\chi$ plane inside which the inequality in Eq.~\eqref{Eq. prior_Im} is satisfied. 

\begin{figure}
    \centering
    \includegraphics[width=\columnwidth,clip=true]{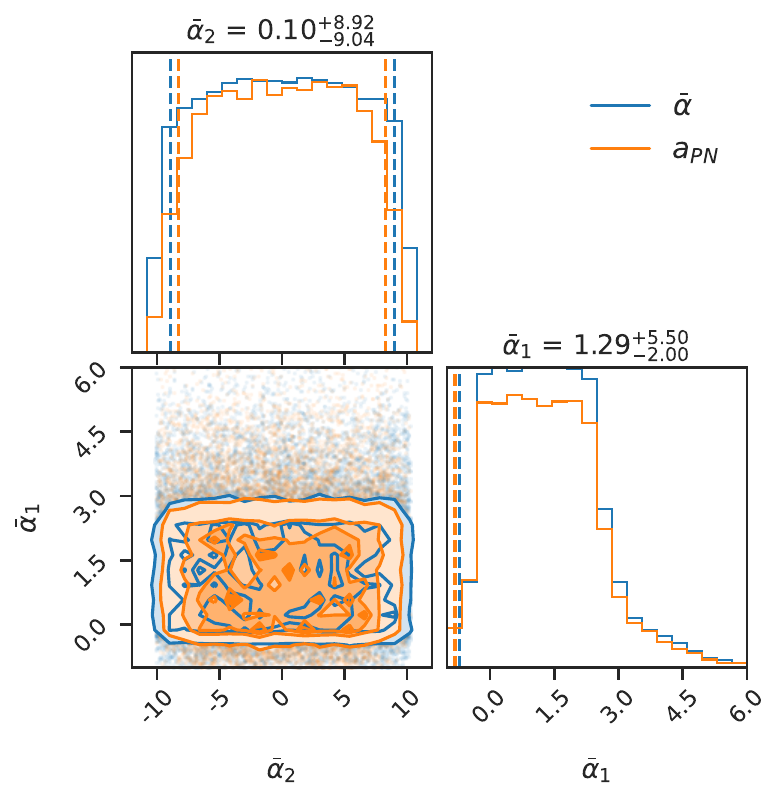}
    \caption{Prior distributions on the $\baq-\baw$ plane. The blue region represents a flat prior on $\baq-\baw$, while the orange region represents a flat prior on $a^{\text{EFT}}_{\text{5PN}} - a^{\text{EFT}}_{\text{6PN}}$, which is then mapped to the $\baq-\baw$ parameters space. Observe that the mapping changes the boundary of the prior in the $\bar{\alpha}_2$ direction, but not in the $\bar{\alpha}_1$ direction.}
    \label{fig.prior}
\end{figure}

Given all of the prior conditions presented above, there are several combinations that one could investigate. One option is to enforce Eqs.~(\ref{Eq. PN_prior_1_ba})-(\ref{Eq. PN_prior_3_ba}) and (\ref{Eq. prior1})-(\ref{Eq. prior_fdamp}) for $\baq$ and $\baw$.
We shall refer to this choice as the $\bar{\alpha}$-prior.
Another option is to enforce Eqs.~(\ref{Eq. PN_prior_1})-(\ref{Eq. PN_prior_3}) and (\ref{Eq. prior1})-(\ref{Eq. prior_fdamp}) for $a^{\text{EFT}}_{\text{5PN}}$ and $a^{\text{EFT}}_{\text{6PN}}$. 
We shall refer to this choice as the $a_{\text{PN}}$-prior.
These different prior choices lead to slightly different prior distributions, which we present in Fig.~\ref{fig.prior}. Observe that the edges of the prior on $\bar{\alpha}_1$ and $\bar{\alpha}_2$ are the same whether we choose to impose the $\bar{\alpha}$-prior or the $a_{\text{PN}}$-prior. The fact that our priors are consistent with each other is re-assuring.

\section{Detectability of EFT Model}
\label{sec.V}

Before carrying out a Bayesian analysis of real GW events, it is worth discussing the detectability of EFT effects with a (synthetic) injection and recovery campaign. The purpose of this section is to show that our model can be distinguished from a GR model (provided the SNR is large tough), and to determine the region of parameter space inside which this is possible. To achieve this, we first extend the \textit{effective cycle} measure proposed by \cite{Sampson:2014qqa} to include amplitude corrections in the waveform model. We then study the recovery of injected signals (with and without GR modifications) using a model that includes GR modifications, through parallel-tempered Markov-Chain Monte-Carlos (PTMCMC) sampling methods, calculating the Bayes factor and the marginalized posterior.

\subsection{Effective Cycles}
\label{sec.V:A}

Assuming Gaussian noise, the likelihood function
\begin{equation}
    \log\ \mathcal{L} = -\frac{1}{2} (h-d|h-d) = -\frac{1}{2}\chi^2
\end{equation}
where $h$ is the waveform model and $d$ is data, and where the inner product is defined as
\begin{equation}
    (a|b) = 4 \Re{\int \frac{\tilde{a}^*(f) \; \tilde{b}(f)}{S_n (f)} df}
\end{equation}
where $\tilde{a}$ and $\tilde{b}$ are the Fourier transforms of $a$ and $b$, and where $S_n (f)$ is the noise spectral density.

Let us now say that hypothesis $\mathcal{H}_0$ is that a signal $d$ can be described by a model $h$, and that hypothesis $\mathcal{H}_1$ is that a signal $d$ can be described by a model $h^\prime$. 
If the data contains the effects described by $h^\prime$, we want to know if we can distinguish this hypothesis from the null hypothesis waveform defined by model $h$. Thus, we set the injected signal to be $h^\prime$, i.e.~$d=h^\prime$, and then, under the Laplace approximation, the logarithm of the Bayes factor is \cite{Cornish:2011ys}
\begin{equation}\label{Eq. BF}
    \begin{aligned}
        \log \ttt{BF} & = \log \frac{e^{-\chi^2 (\mathcal{H}_1)/2}}{e^{-\chi^2 (\mathcal{H}_0)/2}} \frac{\mathcal{O}_1}{\mathcal{O}_0}\,, \\
        & = -\frac{1}{2}(\chi^2 (\mathcal{H}_1) - \chi^2 (\mathcal{H}_0)) + \Delta \log \mathcal{O}\,, \\
        & = \frac{1}{2}\min_{\lambda^a}[(h(\lambda^a)-\hp | h(\lambda^a)-\hp)]  + \Delta \log \mathcal{O}\,, \\
        & \simeq \frac{1}{2} \min_{\lambda^a} [{\chi_{\text{res}}^2}] 
        \simeq \frac{1}{2}\chi^2_{\min}\,,
    \end{aligned}
\end{equation}
where $\lambda^a$ are the parameters of the waveform model $h$, ${\cal{O}}_{1,2}$ is the Occam factor, and $\chi^2({\cal{H}}_{1,2})$ is the inner product between the waveforms in our two hypothesis, while $\chi^2_{\rm res}$ is the inner product of the residual between two waveforms, and $\chi^2_{\rm min}$ is the minimum residual.
In the last line of the above equation, we have dropped the Occam factor because it is subdominant when the fitting factor is close to unity~\cite{Sampson:2014qqa}. 
Using that the signal-to-noise-ratio(SNR) squared is defined as $\rho^2 = (h|h)$ and ${\rho^\prime}^2 = (\hp | \hp)$, we can then write
\begin{equation}\label{Eq. dhdh}
    \begin{aligned}
        \chi^2_{\text{res}} & = (h-\hp | h-\hp) 
                        = \rho^2 + {\rho^\prime}^2 - 2 (h|\hp)\,.
    \end{aligned}
\end{equation}
With this in hand, the logarithm of the Bayes factor is then\begin{equation}\label{Eq. lnBF}
    \ln \text{BF} = \frac{1}{2} \min_{\lambda^a} [ \rho^2(\la) + {\rho^\prime}^2 - 2 (h(\lambda^a)|\hp) ]
\end{equation}

The definition of the fitting factor is
\begin{equation}\label{Eq. FF2}
    \ttt{FF} = \max_{\la} \frac{(h|\hp)}{\sqrt{(\hp | \hp)(h|h)}} = \max_{\la} \left[ \frac{(h(\lambda^a)|\hp)}{\rho(\la) \rho^\prime} \right]\,.
\end{equation}
When $\ttt{FF}\sim 1$, i.e.~when $h$ and $\hp$ are similar to each other, the fitting factor can be written as
\begin{align}
\label{eq:lnBF}
    (1-\ttt{FF}^2) &= \left(1 + \ttt{FF}\right) \left(1 - \ttt{FF}\right) \nonumber \\
    &\simeq 2 \left(1-\max_{\la} \left[ \frac{(h(\lambda^a)|\hp)}{\rho(\la) \rho^\prime} \right] \right)\,.
\end{align}
We can use this to rewrite the Bayes factor as
\begin{equation}\label{Eq. lnBF_FF}
        \ln \ttt{BF}  \simeq \frac{{{\rho^\prime}^2}}{2} \left[ \left(\frac{\rho_{\max}}{{\rho^\prime}}\right)^2 + \left(1-\rtrmrp\right) - \left(\rtrmrp\right)\ttt{FF}^2  \right]\,, 
\end{equation}
where $h_{\max} = h(\lam)$ and $\rho_{\max} = \rho (\lam)$ are the waveform $h$ and the SNR $\rho$ evaluated with the maximum likelihood parameters.

Let us now momentarily assume that we can write the waveforms as $h = A e^{i\Psi}$ and $\hp = A e^{i\Psi^\prime}$, where $(A,\Psi,\Psi')$ are all functions of the GW frequency. Then, when $\rho^\prime = \rho_{\max}$, the Bayes factor becomes \cite{Sampson:2014qqa}
\begin{equation}
    \ln \ttt{BF} \sim \frac{1}{2}\min_{\la} \int \frac{h_c^2 (f) \Delta \Psi^2 (f)}{S_n (f)} d \ln f\,,
\end{equation}
where $h_c = \sqrt{f} A(f)$ is the characteristic strain and $\Delta \Psi$ is the difference between the phase of the two waveforms. With such an expression, it is natural to define the effective cycle via
\begin{equation}
    \mathcal{N}_{\ttt{eff}} = \min_{\delta t, \delta \phi} \left[ \frac{1}{2\pi \rho} \left(  \int  \frac{h_c^2 (f) \Delta \Psi^2 (f)}{S_n (f)} d \ln f  \right)^{1/2}
   \right]
\end{equation}
so that the Bayes factor can be written as
\begin{equation}
\label{Eq:Neff}
    \ln \ttt{BF} \sim 2\pi^2 \rho'{}^2 \min_{\la} \mathcal{N}_{\ttt{eff}}^2
\end{equation}
Thus, with the above expressions for the Bayes factor, the definition of effective cycle appears natural after we align the time and phase of the two waveforms (by minimizing over a time and phase offsets, $\delta t$ and $\delta \phi$).

Given the above expressions, we can easily derive an expression for the effective cycles in terms of the fitting factor. 

Comparing Eq.~\eqref{Eq. lnBF_FF} with~\eqref{Eq:Neff}, we find that
\begin{align}
    \mathcal{N}_{\ttt{eff}} &= \frac{1}{2\pi} \max_{\delta t, \delta \phi} \left[ \left(\frac{\rho_{\max}}{{\rho^\prime}}\right)^2 + \left(1-\rtrmrp\right) 
    \right. \nonumber \\
    &\left. \qquad \qquad \qquad - \left(\rtrmrp\right)\ttt{FF}^2 \right]^{1/2}\,.
\end{align}
When $\ttt{FF} \sim 1$, the effective cycles can be further simplified to
\begin{equation}\label{Eq. Neff_FF}
    \mathcal{N}_{\ttt{eff}} \simeq \frac{1}{2\pi} \max_{\delta t, \delta \phi}  \sqrt{ \left(\frac{\rho_{\max}}{{\rho^\prime}}\right)^2 + 1 - 2\left(\rtrmrp\right)\ttt{FF} }\,.
\end{equation}

Note that in deriving the above two equations we did \textit{not} have to assume that the waveform amplitudes were the same, which thus generalizes the presentation in~\cite{Sampson:2014qqa}.
The effective cycles are thus an ideal tool to estimate the detectability of differences between two waveforms, using computationally inexpensive data analysis tools. 

Let us now represent the waveforms as $h = A e^{i\Psi}$ and $\hp = A^\prime e^{i\Psi^\prime}$, where $(A,A', \Psi,\Psi')$ are all functions of the GW frequency and the amplitudes are not identical. Let us further assume the waveforms are similar to each other, in the sense that $\cos(\Delta \Psi) \simeq 1-\frac{1}{2}(\Delta \Psi)^2$, where $\Delta \Psi = \Psi - \Psi'$. Note that this similarity requirement may not always hold, especially when observing for long time periods and considering deviations that are not so small. When the waveforms are similar to each other, then the fitting factor can be written as
\begin{equation}
        \ttt{FF} \simeq \frac{4}{\rhm \rho^\prime} \int \frac{A^\prime A}{S_n} \left[1-\frac{1}{2}(\Delta \Psi)^2 \right] df\,.
\end{equation}
Inserting this expression into Eq.~(\ref{Eq. Neff_FF}), we have that
\begin{equation}\label{Eq. Neff_exp}
    \begin{aligned}
        \mathcal{N}_{\ttt{eff}} \simeq \frac{1}{2\pi} \max_{\delta t, \delta \phi} & \left[  \left(\frac{\rho_{\max}}{{\rho^\prime}}\right)^2 +1  - \frac{8}{{\rho^\prime}^2} \int \frac{A^\prime A}{S_n}df \right. \\
        & \left. + \frac{4}{{\rho^\prime}^2}\int \frac{A^\prime A}{S_n}(\Delta \Phi)^2 df \right]^{1/2}
    \end{aligned}
\end{equation}
where $A=A(\lam)$ 
and $\Delta\Phi = \Delta \Psi + 2\pi f \delta t - \delta \phi$. When we consider the case in which the amplitudes of $h$ and $\hp$ are the same, Eq.~(\ref{Eq. Neff_exp}) reduces to Eq.~(54) in \cite{Sampson:2014qqa}. To further simplify the above expressions for the effective cycles, we need to make further assumptions about the relation between $h$ and $\hp$, which we do next.

\subsubsection{Assumption $\rhm = \rho^\prime (1+x)$}
\label{sec.V:A.1}

If we assume that $\rhm = \rho^\prime (1+x(\lam))$, for some function $x(\lambda)$, then the Bayes factor in Eq.~(\ref{Eq. lnBF_FF}) becomes
\begin{equation}
    \ln \ttt{BF} \simeq \frac{{{\rho^\prime}^2}}{2} \left[  (1+x)(1-\ttt{FF}^2)  \right]\,,
\end{equation}
where we have neglected terms of $\mathcal{O}(x^2)$.
Since we want to define the effective cycles from its relation with the Bayes factor,
\begin{equation}
    \ln \ttt{BF} \sim 2\pi^2 {\rho^\prime}^2 \min_{\la} [ \mathcal{N}_{\ttt{eff}}^2 ]\,,
\end{equation}
the effective cycle definition of Eq.~(\ref{Eq. lnBF_FF}) becomes
\begin{equation}
    \mathcal{N}_{\ttt{eff}} = \frac{1}{2\pi}  \max_{\delta t, \delta \phi} \left[ \sqrt{(1+x)(1-\ttt{FF}^2)} \right]\,.
\end{equation}
When $\ttt{FF} \sim 1$, the effective cycles further reduce to
\begin{equation}
    \mathcal{N}_{\ttt{eff}} \simeq \frac{1}{2\pi} \max_{\delta t, \delta \phi} \left[ \sqrt{2(1+x)(1-\ttt{FF})} \right]
\end{equation}

The SNR of a waveform, however, only depends on its amplitude, so the assumption $\rhm = \rho^\prime (1+x)$ implies a relation between the two amplitudes, namely
\begin{equation}
        x = \frac{\rhm}{\rho^\prime} - 1 \sim \frac{\delta A}{{A^\prime}}\,.
\end{equation}
This expression means that $x$ should be the same order as ${\delta A}/{{A^\prime}}$, and the effective cycles also receive ${\cal{O}}({\delta A}/{A^\prime})$ corrections from amplitude modifications.
In fact, if we want to expand the fitting factor one step further, it is convenient to start from Eq.~(\ref{Eq. Neff_exp}), so that
\begin{equation}
        \mathcal{N}_{\ttt{eff}} \sim \frac{1}{2\pi} \max_{\delta t, \delta \phi} \left[   x^2 + \frac{4}{{\rho^\prime}^2}\int \frac{A^\prime A}{S_n}(\Delta \Phi)^2 df   \right]^{1/2}\,.
\end{equation}
This expression tells us that the effective cycles only receive ${\cal{O}}(x^2)$ corrections from pure amplitude modifications, as we will see more clearly in the next discussion.

\subsubsection{Assumption $A(\lam)=A^\prime + \delta A$}
\label{sec.V:A.2}

Under this assumption, to simplify the expression of the effective cycles, we still need to derive a relation between $\rhm$ and $\rho$. Using the definition of $\rhm$, we have that
\begin{equation}
    \begin{aligned}
        \rhm^2 & = 4 \int \frac{A^2(\lam)}{S_n} df \\
               & = {\rho^\prime}^2 (1 + 2y + y^{(2)})
    \end{aligned}
\end{equation}
where we defined 
\begin{align}
y=\frac{4}{{\rho^\prime}^2} \int \frac{A^\prime \delta A}{S_n}df     
\end{align}
and
\begin{align}
y^{(2)} = \frac{4}{{\rho^\prime}^2} \int \frac{(\delta A)^2}{S_n}df\,,    
\end{align}
so that $\rhm = \rho^\prime \sqrt{1+2y+y^{(2)}}$. Then, the effective cycles of Eq.~(\ref{Eq. Neff_exp}) become
\begin{equation}
    \begin{aligned}
        \mathcal{N}_{\ttt{eff}} \simeq \frac{1}{2\pi} & \max_{\delta t, \delta \phi} \left[   y^{(2)} 
        + \frac{4}{{\rho^\prime}^2}\int \frac{{A^\prime}^2}{S_n}(\Delta \Phi)^2 df
        \right. \\
                                & \left. \qquad + \frac{4}{{\rho^\prime}^2}\int \frac{A^\prime\delta A}{S_n}(\Delta \Phi)^2 df \right]^{1/2}                          
    \end{aligned}
\end{equation}
The $y^{(2)}$ quantity is the quadratic part of the amplitude difference,
\begin{align}
    y^{(2)} & = \frac{4}{{\rho^\prime}^2} \int \frac{(\delta A)^2}{S_n}df = \frac{\int \frac{(\delta A)^2}{S_n}df}{\int \frac{{A^\prime}^2}{S_n}df} \sim \left(\frac{\delta A}{A^\prime}\right)^2\,.
\end{align}
Another part of the contribution coming from the amplitude is
\begin{align}
    \delta \mathcal{N} & = \frac{4}{{\rho^\prime}^2}\int \frac{A^\prime\delta A}{S_n}(\Delta \Phi)^2 df = \frac{\int \frac{A^\prime\delta A}{S_n}(\Delta \Phi)^2 df}{\int \frac{{A^\prime}^2}{S_n}df} 
    \nonumber \\
    &\sim \frac{\delta A}{A^\prime}(\Delta \Phi)^2
\end{align}
Clearly, whether we can neglect the $y^{(2)}$ term depends on whether $({\delta A}/{A^\prime})$ is of the same order as $(\Delta \Phi)^2$ or not.

\begin{figure}[h!]
    \centering
    \includegraphics[width=\columnwidth]{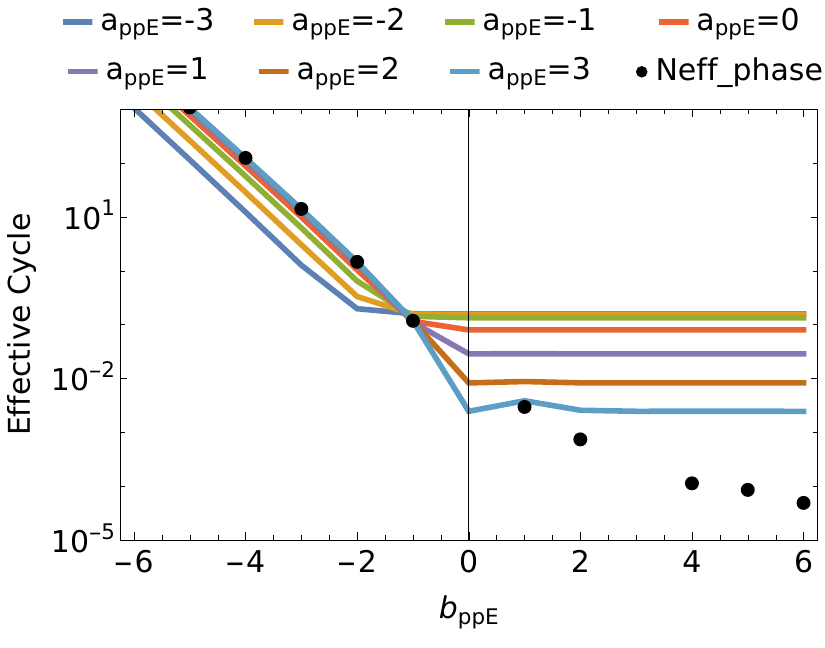}
    \caption{Effective cycles of modified waveforms with different amplitude and phase corrections as a function of the ppE exponent parameters. Different colors denote different ppE exponent parameters in the GW amplitude modification, and black dots stand for effective cycles without amplitude corrections. Observe that when $b_{\rm ppE}>0$, amplitude corrections should be included in the effective cycle calculation.}
    \label{fig.Neff}
\end{figure}

To assess the influence of including amplitude corrections to the effective cycle calculation, we take the ppE waveforms~\cite{Yunes:2009ke} 
\begin{equation}
    \tilde{h}(f) = \tilde{h}_{\text{GR}}(1+\alpha_{\text{ppE}} u^{a_{\text{ppE}}})e^{i\beta_{\text{ppE}} u^{b_{\text{ppE}}}}
\end{equation}
as a proxy, where $\tilde{h}_{\text{GR}}$ is the GR Fourier waveform, $(\alpha_{\text{ppE}},\beta_{\text{ppE}})$ are ppE amplitude parameters, and $(a_{\text{ppE}},b_{\text{ppE}})$ are ppE exponent parameters. Figure~\ref{fig.Neff} presents a rough estimate of the effect of phase and amplitude corrections in tests of GR through an effective cycle calculation, where we set $\alpha_{\text{ppE}}=\beta_{\text{ppE}}=1$. Observe that when the ppE exponent of the phase correction $b_{\text{ppE}}$ is negative, the effective cycles are dominated by the phase correction, and the amplitude corrections can be neglected. However, when $b_{\text{ppE}}$ becomes positive, phase corrections stop affecting the effective cycles as much as amplitude corrections. Since the EFT corrections we study in this paper appear as positive ppE exponents in the phase correction, we expect that amplitude corrections will have to be included, as we do in this paper. 

\subsection{Injection and Recovery}
\label{sec.V:B}

Now that we have a rough understanding of the importance of amplitude corrections, let us carry out a synthetic injection and recovery campaign to determine the regime of parameter space inside which EFT modifications can be constrained by current GW data. We will consider two types of injections:
\begin{itemize}
    \item \textbf{GR Injections}. We inject with an IMRPhenomD model, without any GR deviations (i.e.~with the EFT-corrected IMRPhenomD model but with $\bar{\alpha}_1 = 0 = \bar{\alpha}_2$. 
    \item \textbf{EFT Injections}. We inject with the EFT-corrected IMRPhenomD model, setting $(\baq, \baw)=(2,3)$ to study relative small EFT effects, and $(\baq,\baw)=(3,4)$ to study relatively large deviations.
\end{itemize}
In all the above cases, we inject with masses $m_1=15 M_\odot$ and $m_2 = 10 M_\odot$, and with dimensionless spins $\chi_1 = 0.2$ and $\chi_2 = 0.1$.

Given these injections, we will then recover them with the EFT-corrected IMRPhenomD model, whose parameters are $\vec{\lambda}= \{ m_{1,2}, \chi_{1,2}, \psi, \text{ra}, \text{dec}, D_L, \iota, t_c, \phi_c, \baq, \baw \}$. The priors on these parameters are as follows. We choose flat priors for the component masses, spin magnitude, and the volume for the luminosity distance.
Moreover, we choose an isotropic prior for spin and binary orientation, and we assume isotropic distributions on the sky map. For the EFT coupling parameters, we explore both a flat prior for $\bar{\alpha}$ and the $a_{\rm PN}$ prior discussed in Sec.~\ref{sec.IV}. The choice of exploring a flat prior is to guarantee that the posteriors we obtain are not driven by the non-trivial $a_{\rm PN}$ prior. 

Given this model with the priors discussed above, and the injections listed above, we then explore the $13$-dimensional likelihood function using a PTMCMC sampling algorithm. In particular, we use the \textit{Gravitational Wave Analysis Tools} (GWAT) code, developed in~\cite{Perkins:2021mhb}, with the following options. We choose $16$ temperatures and $8$ chains for each temperature. With adaptive temperature ladders \cite{Vousden:2016eeu} and the mixing proposals including Gaussian proposal, differential evolution, Fisher matrix and GMM proposal, we obtain about $40,000$ independent samples for each Bayesian analysis.  
Before analyzing the posterior distributions obtained from this PTMCMC sampling, we carried out several tests to ensure the chains had properly converged. First, we eliminate a burn-in period of $10,000$ iterations, which is determined by investigate the trace plots. We then ran the analysis several times with different seeds (i.e.~different starting points in the parameter space) and checked the stability of the trace plots to ensure the chains had converged.

\begin{figure*}[htb]
    \centering
    \includegraphics[width=\columnwidth,clip=true]{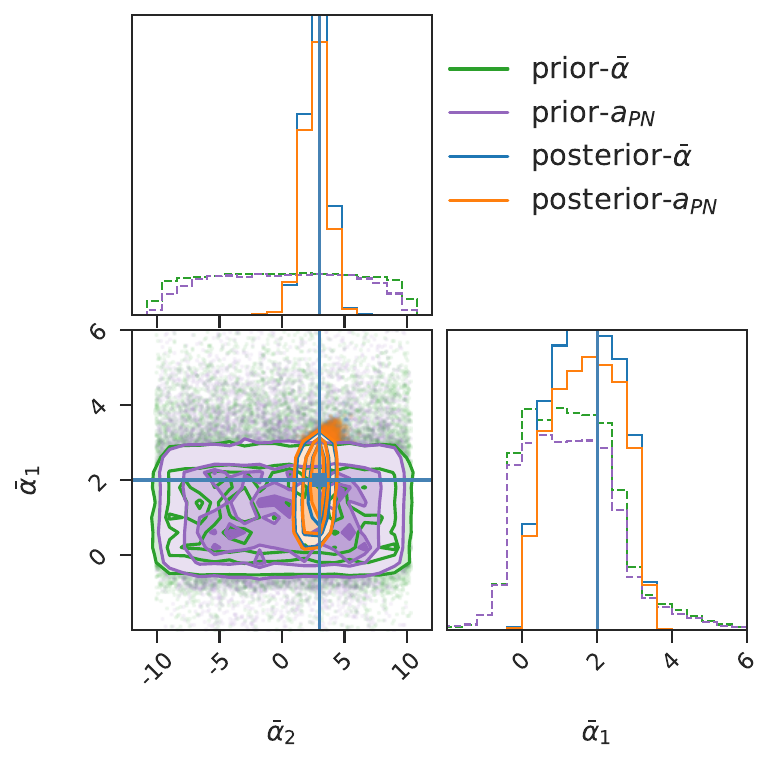}
    \quad
    \includegraphics[width=\columnwidth,clip=true]
    {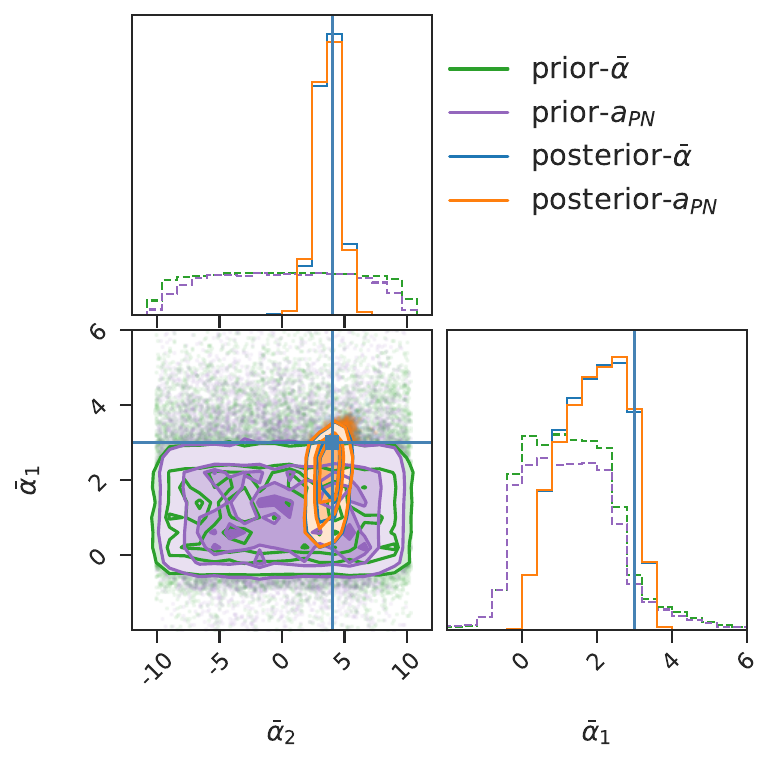}
    \caption{Two-dimensional posterior corner plots of the EFT parameter space for small (left panel) and large (right panel) EFT injections (with vertical solid lines indicating the injected values). The marginalized posterior distributions (solid lines) are compared to the priors (dashed lines) when we use the $\bar{\alpha}$ prior choice (green and blue) and the $a_{\rm PN}$ prior choice (purple and orange). Observe that in all cases, the posteriors are informative for the $\bar{\alpha}_2$ parameter, but they are informative for the $\bar{\alpha}_1$ parameter only for large EFT deviations. 
    }
    \label{fig.injection_53}
\end{figure*}

Let us first discuss our results when we inject with EFT signals. In Fig.~\ref{fig.injection_53}, we compare the posteriors obtained when analyzing injections with small (left) and large (right) EFT corrections, using the two sets of priors on the EFT coupling parameters discussed earlier. Observe that the posteriors are consistent with each other when using these two choices of priors, indicating that the likelihood is dominating the parameter estimation. Observe also that all marginalized posteriors are consistent with the injected values. The measurement of $\baw$ is more accurate than that of $\bar{\alpha}_1$ because the former enters at 5PN order, and thus it has a stronger effect in the inspiral. Finally, observe that the posteriors are informative because they are always different from the priors, except in the $\bar{\alpha}_1$ case for injections with small EFT corrections. 
The results presented above indicate that, given a sufficiently loud EFT injection, we can both extract it with the EFT-corrected IMRPhenomD model and distinguish it from a GR model. Put another way, the injections we considered are informative enough to prefer the EFT-corrected IMRPhenomD model over a GR IMRPhenomD model, as quantitatively demostrated by the log evidences shown in Table~\ref{tab.evidence}. The difference of the log evidences corresponds to the log Bayes factor, which is always in favor of the non-GR model. 

\begin{table}
\centering
    \begin{tabular}{|c|c|c|c|}
    \hline
       \backslashbox{Signal}{Model}  & GR & EFT($\bar{\alpha}$) & EFT(PN)\\ 
    \hline
        GR                           & 149.64 & 146.57 & 146.52 \\
    \hline
        EFT($\bar{\alpha}_1=2, \bar{\alpha}_2=3$)                         & 141.42 & 144.45 & 144.56 \\
    \hline
        EFT($\bar{\alpha}_1=3, \bar{\alpha}_2=4$)                         & 132.64 & 142.63 & 142.71 \\
    \hline
    \end{tabular}
    \caption{Log evidence of the Bayes inference with different waveform models and priors on different injections. Since the Bayes factor is the ratio of the evidences, the log Bayes factor can be calculated by eye by taking the differences of the log evidences presented in this table. In particular, observe that the EFT model is strongly preferred over the GR model for both EFT injections, regardless of the priors used.}
    \label{tab.evidence}
\end{table}

We have also carried out an injection-recovery study using a GR injection. This analysis reveals how well we can constrain EFT parameters around zero, given a signal that is consistent with GR. Instead of presenting these results here, however, we shall present them in the next section, where we also analyze real GW data. By comparing GR injections with GW data, we will be able to further establish that the observed data is indeed consistent with GR. 

\section{Bayesian parameter estimation study to probe EFT modifications with real GW data}
\label{sec.VI}

Let us now use the EFT-corrected IMRPhenomD model to analyze GW data, focusing in particular on the GW170608 event~\cite{LIGOScientific:2017vox}. This event was detected by the advanced LIGO observatories and, when analyzed with GR waveform models, it was inferred to have been produced by the coalescence of two black holes with masses of $\sim 12 M_\odot$ and $\sim 7 M_\odot$ at $\sim 340$ Mpc from Earth, assuming a GR waveform model.

\begin{figure}
    \centering
    \includegraphics[width=\columnwidth]{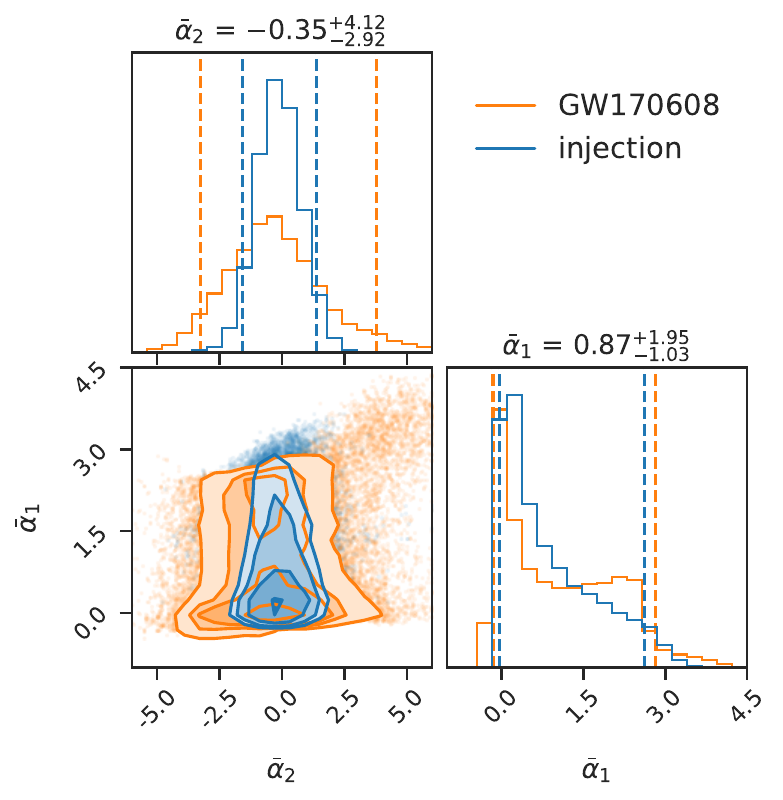}
    \caption{Corner posterior plot of the $\bar{\alpha}_1$-$\bar{\alpha}_2$ subspace obtained by analyzing the GW170608 event (orange) and a similar GR injection (blue), using the EFT-corrected IMRPhenomD model, together with their 90\% confidence limits (dashed vertical lines). Observe that the analysis of the GW170608 event and the GR injection lead to very similar posteriors. Observe also that the marginalized posterior distribution places stringent constraints on the EFT coupling parameters, namely 
    $-0.16 < \bar{\alpha}_1 < 2.82$ and $-3.27 < \bar{\alpha}_2 < 3.77$
    at 90\% confidence. }
    \label{fig.posterior_170608_inj00}
\end{figure}

We now analyze this GW event with the EFT-corrected IMRPhenomD model and the same priors discussed in the previous section. The GW data and the power spectral density during the event are both publicly available, and we take them from GWOSC~\cite{KAGRA:2023pio, LIGOScientific:2019lzm}. 
The posteriors we obtain are shown in Fig.~\ref{fig.posterior_170608_inj00}, which we observe are consistent with those obtained when analyzing the GR injection, implying that the data prefers GR over an EFT description of the signal. This is indeed verified by our calculation of the Bayes factor, 
$\log \mathcal{B}^{\text{EFT}}_{\text{GR}} = -2.81$, which indicates a clear strong preference of the data for the GR model. Observe also that the posteriors are informative, as they are very different from the priors (compare to Fig.~\ref{fig.prior}). Finally, observe that the posteriors allow us to place a 90\% confidence limit on the EFT coupling constants, bounding 
$-0.16 < \bar{\alpha}_1 < 2.82$ and $-3.27 < \bar{\alpha}_2 < 3.77$
to 90\% confidence. These constraints are $3.5$ times stronger than other previous constraints, because the EFT-corrected IMRPhenomD model we developed and deployed on the data is more informative (containing both inspiral and ringdown modifications) than any other model previously used. In particular, observe that the constraints are asymmetric, strongly restricting the $\bar{\alpha}_1 < 0$ region. This is because for such negative values, the EFT model predicts QNMs that are exponentially growing (and thus, imply unstable BHs).  

\section{Conclusions and Discussion}
\label{sec.VII}

We have here constructed a model for the GWs emitted by the quasi-circular inspiral, merger and ringdown of spin-aligned/anti-aligned BHs in cubic EFT, thus extended the IMRPheonmD model to an EFT-corrected IMRPhenomD model. We incorporated the EFT corrections to the inspiral phase  amplitude (to first sub-leading order in the PN approximation), as well as the EFT corrections to the QNM frequencies (using a fit valid for BHs with moderate spins), to generate a full description of EFT effects. Furthermore, we investigated the model we created and explained its limitations, which restrict the region of EFT parameter space where the model is valid (e.g. to avoid QNMs that are exponentially growing instead of decaying). 

With this model in hand, we then investigated whether the EFT corrections the model encodes are measurable with GW observations. We first extended the concept of effective cycles to account for waveforms that include both modified amplitudes and phases. We found that for theories of gravity that introduce positive PN order corrections to the phase, the inclusion of amplitude corrections is very important to improve the distinguishability of the model relative to GR. We then carried out a synthetic injection and recovery campaign to determine (i) whether the EFT-corrected IMRPhenomD model can detect non-GR signals, and (ii) whether constraints on the EFT coupling parameters can be placed when analyzing GR signals, finding that this is so in both cases. Finally, we analyzed real GW data, the GW170608 event, and found that the data prefers GR over the EFT model, yielding new constraints on the EFT coupling parameters. These constraints are $3.5$ times stronger than previous constraints because the EFT model we developed and deployed on the data is more informative than previous models used (as it includes both modifications to the inspiral and ringdown stage of coalescence). 

Our work points to several directions for possible extensions. One such direction is to extend the PN approximation of the inspiral stage to higher order. The calculations we presented here are valid to first sub-leading PN order, and thus, they are of 5 and 6PN order relative to the leading PN order term in GR. Another direction for future work is to calculate the QNM frequencies at higher values of the remnant BH spin, and to calculate more accurate fits for the ringdown stage. Both of these extensions of our work should increase the robustness of our results, although it probably will not modify the bounds appreciably. 

A third possible direction one could pursue is the modeling of the merger stage itself in cubic EFT theories. The IMRPhenomD model in GR is calibrated in the ``intermediate stage'' to numerical relativity simulations of mergers in GR. The same is not possible in cubic EFT theories because such numerical relativity simulations do not yet exist. The EFT-corrected model that we developed, therefore, does not contain direct modifications to the intermediate stage; this stage is indeed EFT modified, but only because the inspiral stage is modified and the intermediate-inspiral boundary must be continuous and differentiable. Numerical relativity simulations in cubic EFT theories would allow us to redo the intermediate stage fits, to then develop a more complete EFT-corrected IMRPhenomD model.

\section{Acknowledgement}
The authors thank Rohit S.~Chandramouli for helping us clarify the extension of effective cycles for theories with amplitude modifications. We also thank Hector O.~Silva for clarifying some results related to QNMs in cubic EFT theory.  H.-Y. L. is supported by the Joint PhD Training program from University of Chinese Academy of Sciences. N.~Y.~is support from the Simons Foundation through Award No. 896696, the National Science Foundation (NSF) Grant No. PHY-2207650 and NASA through Grant No.~80NSSC22K0806. 

\appendix

\section{Calculation of EFT-corrected Waveform}
\label{sec:appendix}

In this Appendix, we derive the modified version of Kepler's third law given the EFT action considered in this paper to 1PN order. Then, we calculate the EFT corrected energy and flux with such a modified Kepler law also to 1PN order. Given the energy and flux, we finally obtain the EFT waveform in the SPA method to 1PN order. Since we are carrying out a PN calculation, in this appendix we restore the factors of Newton's gravitational constant $G$ and the speed of light $c$ to make the PN order counting easier to recognize.

\subsection{Modified Kepler relation}

Reference \cite{Brandhuber:2019qpg,Emond:2019crr} derived the EFT correction to the gravitational potential. Including the 1PN order part of the GR sector in the gravitational potential, we have
\begin{equation}\label{Eq. Vrp_1PN}
\begin{aligned}
     V(\vec{r}, |\vec{p}|) = & - \frac{G \eta M^2}{r}  -  \frac{\eta M}{8}(1-3\eta) \frac{v^4}{c^2} \\
     & + \frac{\eta M}{c^2} \left[ -\frac{GM}{2r}(3+\eta) v^2 + \frac{(GM)^2}{2 r^2}\right] \\ 
      & + {\cal{O}}(1/c^4) + \frac{3}{8} \frac{\alpha_1 G^2}{r^6 c^{10}} \frac{M}{\eta} \frac{{p}^2}{c^2} \\
      & - \frac{3}{4} \frac{\alpha_2 G^2}{r^6 c^{10}} \eta M^3 \left(1 - \frac{m_1^2 + m_2 ^2}{2 \eta^2 M^4} \frac{{p}^2}{c^2}\right)\\
      & {\cal{O}}(\alpha_i/c^{14},\alpha_i^2)\,,
\end{aligned}
\end{equation}
where $v$ is the velocity in the central-of-mass frame, $r$ is the orbital separation and $\vec{p}$ is the linear 3-momentum in the central-of-mass frame, with $p^2$ the square of its magnitude (computed with the flat metric). Since the potential not only contains $p^2$ terms but also $v^4$ terms, we use that $p=\mu v$ and also $v = r\Omega$ by definition. 

To convert potential $V(r,\vec{p})$ into one that depends only on the orbital separation, $V(r)$, we need to first find a relation between the momentum $\vec{p}$, the orbit separation $r$, and $\Omega$. Noting that $\vec{p}$ only appears in the EFT terms, we only need to find the relation $\vec{p}(r,\Omega)$ in GR, if we are only interested in the $\mathcal{O} (\bar{\alpha})$ corrections. Moreover, since we are considering quasi-circular orbit, $p_r = 0$ and $p^2 = p_r^2 + p_\phi^2 / r^2 = p_\phi^2 / r^2 $. Given this, the $phi$ component of momentum is defined as usual,
\begin{align}
\mL_\text{GR} &=  \frac{1}{2}\mu v^2 + \frac{G \eta M^2}{r}  -  \frac{\eta M}{8c^2}(1-3\eta) v^4 
\nonumber \\
& - \frac{\eta M}{c^2} \left[ -\frac{GM}{2r}(3+\eta) v^2 + \frac{(GM)^2}{2 r^2}\right]
\nonumber \\
&+ {\cal{O}}(1/c^4)\,, 
\end{align}
we find
\begin{equation}
    p_\phi = \mu r^2 \Omega \left[ 1 + \left(   3 + \eta  \frac{1}{2} - \frac{3}{2}\eta   \right) \frac{GM}{c^2 r} + {\cal{O}}(1/c^4) \right]\,.
\end{equation}

With this expression, we can rewrite the potential in Eq.~\eqref{Eq. Vrp_1PN} entirely as a function of $r$ and $\Omega$.

With this in hand, we then derive the modified version of Kepler's third law by using the Euler-Lagrange equation
\begin{equation}
    \frac{d}{dt}\frac{\partial \mc{L}}{\partial \dot{r}} - \frac{\partial \mc{L}}{\partial r} = 0\,.
\end{equation}
In particular, since we are considering quasi-circular orbits, the Euler-Lagrange equations becomes simply
\begin{equation}\label{Eq. EL_EQ}
    \frac{\partial \mc{L}}{\partial r} = 0\,,
\end{equation}
with the Lagrangian
\begin{equation}
    \mc{L} = \frac{1}{2}\mu v^2 - V\,,
\end{equation}
as explained e.g.~in~\cite{Blanchet:2013haa}.

Equation~\eqref{Eq. EL_EQ} then becomes an algebraic equation for $\Omega$ as a function of $r$, which one can solve perturbatively in a PN expansion. 
Defining the PN variable $\gamma = GM/c^2 r$, the EFT corrected version of Kepler's third law becomes
\begin{equation}\label{Eq. Omega2_n}
    \Omega^2 = (\Omega^2)_\text{GR} + \frac{c^8 \alpha_1}{(GM\Lambda)^4} (\Omega^2)_{\alpha_1} + \frac{c^8 \alpha_2}{(GM\Lambda)^4} (\Omega^2)_{\alpha_2} + {\cal{O}}(\alpha_i^2)\,.
\end{equation}
where
\begin{equation}\label{Eq. Kepler_law_EFT}
    (\Omega^2)_\text{GR} = \frac{\gamma^3}{(GM)^2} \left[1 + (\eta-3) \gamma + {\cal{O}}(1/c^4) \right]
\end{equation}
is the standard GR result at 1PN order, while
\begin{align}
    (\Omega^2)_{\alpha_1} & = -\frac{3}{2 (GM)^2} \gamma^9 \left[1 +  \left(\frac{9}{2}\eta-\frac{5}{2}\right)\gamma + {\cal{O}}(1/c^4)\right] \\
    \label{Eq. Omega_2_alpha2}(\Omega^2)_{\alpha_2} & = \frac{9}{2 (GM)^2} \gamma^8 \left[1 + \left(\frac{19}{6}\eta-\frac{17}{6}\right) \gamma + {\cal{O}}(1/c^4)\right] 
\end{align}
are the EFT corrections to 1PN order. These corrections to Kepler's third law was also derived in \cite{AccettulliHuber:2020dal}, but the result was only correct to leading PN order; we here extended these results to 1PN order, including the 1PN contribution in GR times the 5PN contribution due to $\bar{\alpha}_2$.

In PN theory, one finds it convenient to also derive an expression for the inverse relation between $\Omega$ and $r$. Defining the PN variable $x=(GM\Omega/c^3)^{2/3}$, we then find that 
\begin{equation}\label{Eq. gamma_x}
    \gamma = \gamma_{GR} + \frac{\alpha_1}{(GM\Lambda)^4} \gamma_{\alpha_1} + \frac{\alpha_2}{(GM\Lambda)^4} \gamma_{\alpha_2}\,,
\end{equation}
where the GR contribution is
\begin{align}
    \gamma_{GR} & = x + \left(1-\frac{\eta}{3}\right)x^2\,,
\end{align}
while the EFT corrections are
\begin{align}
    \gamma_{\alpha_1} & = \frac{x^7}{2} \left[ 1 + \left( \frac{81 + 5 \eta}{24}\right) x + {\cal{O}}(1/c^4) \right] \\
    \gamma_{\alpha_2} & = -\frac{3}{2}x^6 \left[  1 - \left( \frac{\eta-43}{6} \right) x + {\cal{O}}(1/c^4) \right]
\end{align}
Finally, one can also find a relation for the radius $r$ as a function of the variable $x$, 
namely 
\begin{equation}
    r=r_{GR} + \frac{c^8 \alpha_1}{(GM\Lambda)^4} r_{\alpha_1} + \frac{c^8 \alpha_2}{(GM\Lambda)^4} r_{\alpha_2}\,,
\end{equation}
where the GR expression is
\begin{align}
    r_\ttt{GR} & = \frac{GM}{x} + \left(  \frac{\eta}{3} - 1 \right) GM + \mathcal{O}(x)\,,
\end{align}
and the EFT corrections are
\begin{align}
    \label{Eq. r_EFT_a1} r_{\alpha_1} & = -GM x^5 + \mathcal{O}(x^6) \\
    \label{Eq. r_EFT_a2} r_{\alpha_2} & = 3 GM x^4 + \frac{31+3\eta}{4 }GM x^5 + \mathcal{O}(x^6)\,.
\end{align}

Let us now compare all of the above results with those that have already appeared in the literature. If we define the PN variable $v_f = (\pi G M f/c^3)^{1/3} = (GM\omega/2c^3)^{1/3} = (GM\Omega/c^3)^{1/3} = x^{1/2}$, the results in Eqs.~\eqref{Eq. r_EFT_a1} and \eqref{Eq. r_EFT_a2} coincide with Eq.~(4.2) in \cite{AccettulliHuber:2020dal}, except for the 1PN order part of $\alpha_2$. As we mentioned, the discrepancy is due to the fact that \cite{AccettulliHuber:2020dal} did not include the 1PN part of the GR sector in their calculations. We find that these terms are important to obtain a 1PN accurate inspiral model in the EFT sector.

\subsection{EFT correction to energy and flux}

With this in hand, it is straightforward to find the orbital energy, using that 
\begin{equation}
\begin{aligned}
    E & = H/M = \frac{1}{M}\left(\frac{\partial \mc{L}}{\partial \dot{\phi}}\dot{\phi} - \mc{L}\right) \\
    &= \frac{1}{M}\left(\frac{\partial \mc{L}}{\partial (r \dot{\phi})}(r\dot{\phi}) - \mc{L}\right) = \frac{1}{M}\left(\frac{\partial \mc{L}}{\partial v}(v) - \mc{L}\right)
\end{aligned}
\end{equation}
Before we can do so, however, we must first find a way to write $v$ as a function of $r$. The energy depends on $v^2$, which satisfies 
$v^2 = r^2 \Omega^2 = {(GM\Omega/c^3)^2}/{\gamma^2}$.
Inserting Eq.~\eqref{Eq. Omega2_n} into the above equation, $v$ can be written as the following series in $\gamma$ 
\begin{align}
    v^2 &= \gamma + (\eta - 3)\gamma^2 + {\cal{O}}(\gamma^3)
    \nonumber \\
    & \frac{9\bar{\alpha}_2}{2} \gamma^6 - 3\frac{2\baq + (17-19\eta) \baw}{4}\gamma^7 + \mathcal{O}(\alpha_i \gamma^8)
\end{align}
Then, using Eq.~\eqref{Eq. gamma_x} to replace $\gamma$ with $x$, we can find an expression for $v$ in terms of $x$, namely
\begin{align}
    v^2 &= x + \left(  \frac{2\eta}{3} - 2  \right) x^2 + {\cal{O}}(x^3) 
    \nonumber \\
    & + 3 \; \baw \ x^6 + \left( -\baq + 5\frac{5+\eta}{}\baw   \right)x^7 + \mathcal{O}(x^8)\,.
\end{align}

Finally, putting all of this together, the energy is
\begin{equation}\label{eq.Energy_EFT}
    E = E_\text{GR} + \alpha_1 E_{\alpha_1} + \alpha_2 E_{\alpha_2} + {\cal{O}}(\alpha_i^2)
\end{equation}
where 
\begin{equation}\label{eq.E_tot}
    E = E_\text{GR} + \frac{c^8\alpha_1}{(GM\Lambda)^4} E_{\alpha_1} + \frac{c^8\alpha_2}{(GM\Lambda)^4} E_{\alpha_2} + {\cal{O}}(\alpha_i^2)
\end{equation}
with
\begin{align}
    \label{eq.E_GR_1PN}E_\text{GR} & = -\frac{1}{2}\eta x + \frac{\eta}{24}(9+\eta)x^2 + \mathcal{O}(x^3) \\
    \label{eq.E_a1}E_{\alpha_1} & = -11\eta x^7 + \mathcal{O}(x^8)\\
    \label{eq.E_a2}E_{\alpha_2} & = \eta \left[   \frac{9}{4}x^6 +  \frac{121-22\eta}{8}x^7 + \mathcal{O}(x^8)  \right]\,.
\end{align}
Again, we can replace $x$ with the PN variable $v_f$ and compare the energy we obtain with Eq.~(4.6) in \cite{AccettulliHuber:2020dal}. We find that the leading PN order parts that are proportional to $\alpha_1$ and $\alpha_2$ are in agreement. The difference is that we have included the 1PN order part of the GR term, which corrects the next PN order term of $\alpha_2$.

Reference \cite{AccettulliHuber:2020dal} already derived the EFT correction to the flux in their Eq.~(3.17), except that they pulled out the Newtonian quadrupole moment $Q^{ij}_N$ as a prefactor. We will not do this here because we wish to also consider the 1PN order terms in the GR sector. Thus, we write the flux as
\begin{equation}\label{eq.Flux_EFT_old}
    \mc{F} = \mc{F}_\text{R} + \mc{F}_{I_1} + \mc{F}_{I_2}
\end{equation}
where $\mc{F}_\text{R}$ corresponds to the contribution from Ricci scalar term in the action, while $\mc{F}_{I_1}$ and $\mc{F}_{I_2}$ correspond to the contributions from the $I_1$ and $I_2$ terms (see Eq. (\ref{Eq. L_D6})). For $\mc{F}_{I_1}$ and $\mc{F}_{I_2}$, we use the results of \cite{AccettulliHuber:2020dal}, where they derive these fluxes by comparing the quadrupole moment terms in the effective point action with scattering amplitude calculations. However, if we write the flux in form of
\begin{equation}\label{eq.Flux_EFT}
    \mc{F} = \mc{F}_\text{GR} + \baq \mc{F}_{\alpha_1} + \baw \mc{F}_{\alpha_2}
\end{equation}
where $\mc{F}_\text{GR}$ is the GR flux, then
\begin{equation}\label{eq.FF_FF}
     \baq \mc{F}_{\alpha_1} + \baw \mc{F}_{\alpha_2} = \mc{F}_{I_1} + \mc{F}_{I_2} + \bar{\mc{F}}_\text{R}\,,
\end{equation}
where $\bar{\mc{F}}_\text{R}$ is the modifications that come from the GR flux evaluated with the modified Kepler relation. 
To obtain the full 1PN result, we would need to include higher moments into the effective point particle action, and then calculate scattering amplitudes contracted with the corresponding terms, as done for the quadrupole moment \cite{AccettulliHuber:2020dal}, all of which is well beyond the scope of this work. Thus, we will assume here that the correction coming from higher PN order moments to the EFT effects in the flux is small, which means that we will use the result from \cite{AccettulliHuber:2020dal} for $\mc{F}_{I_1} $ and $ \mc{F}_{I_2}$ and the Newtonian form of the flux for the last term in Eq.~\eqref{eq.FF_FF}. Then, we will directly add the 1PN term for $\mc{F}_{GR}$ into Eq.~\eqref{eq.Flux_EFT}, since we still need it in the GW phase and amplitude calculation.

Putting all of this together, the flux is
\begin{equation}\label{eq.Flux_tot}
    \mc{F} = \mc{F}_\text{GR} + \alpha_1 \mc{F}_{\alpha_1} + \alpha_2 \mc{F}_{\alpha_2}  + {\cal{O}}(\alpha_i^2)\,,
\end{equation}
where
\begin{align}
    \label{eq.Flux_GR_1PN}\mc{F}_\text{GR} & = \frac{32c^5}{5G}\eta^2 v^{10} \left[   1+\left(-\frac{1247}{336} - \frac{35}{12}\eta\right) v^2  + {\cal{O}}(1/c^3)  \right] \\
    \label{eq.Flux_a1}\mc{F}_{\alpha_1} & = -\frac{32c^5}{5G}\eta^2 v^{10} \left[ \frac{384}{(GM)^4}v^{12} + {\cal{O}}(1/c^{13}) \right] \\
    \label{eq.Flux_a2}\mc{F}_{\alpha_2} & = \frac{32c^5}{5G}\eta^2 v^{10} \left[    \frac{144}{(GM)^4}v^{10} + 84\frac{1+\eta}{(GM)^4}v^{12}  + {\cal{O}}(1/c^{13})  \right]\,.
\end{align}

One can also verify that the leading PN order part of the EFT terms coincides with Eq.~(4.11) in \cite{AccettulliHuber:2020dal}.

\subsection{Waveform}

The stationary Phase Approximation (SPA) assumes that the dominant contribution to the integral
\begin{equation}
    \tilde{h}(f) = \int h(t) e^{i \psi(t)} dt
\end{equation}
comes from the regime near the stationary point $t_f$, defined by
\begin{equation}
    \dot{\psi}(t_f) = 0\,.
\end{equation}
Expanding the amplitude and phase of the integrand around $t_f$, and replacing the derivative with respect to time with $\mc{F} = - \dot{E}$ through the use of the chainrule,  the final result of the SPA method is
\begin{equation}
    \tilde{h}^{\rm SPA}(f) = \frac{A(t_f)}{\sqrt{\dot{f}(t_f)}} e^{i\psi_f - \frac{\pi}{4}}\,.
\end{equation}
where the Fourier phase is
\begin{equation}\label{Eq. phase_SPA_2}
    \psi_f = 2\pi f t_{\text{ref}} - \phi_{\text{ref}} + \int^{v_{\text{ref}}}_{v_f} (v_f^3 - v^3)\frac{E^\prime (v)}{\mathcal{F}(v)} dv\,,
\end{equation}
which was already presented in Eq.~\ref{Eq. phase_SPA}.

Let first focus on the amplitude of the Fourier transform in the SPA. We substitute the derivative with respect to time of the frequency, using that $\mc{F} = - \dot{E}$ and the chain rule, to obtain
\begin{equation}\label{eq.SPA_amp}
    \frac{A(t_f)}{\sqrt{\dot{f}(t_f)}} = \frac{4}{c^5 D_L} \eta GM (r^2 \Omega^2) \frac{M}{v_f} \sqrt{-G \frac{\pi}{3}\frac{E^\prime}{\mc{F}}}
\end{equation}
Evaluating this expression by inserting Eqs.~(\ref{eq.E_tot})-(\ref{eq.E_a2}) and~(\ref{eq.Flux_tot})-(\ref{eq.Flux_a2}), we find that the total amplitude is $A(f) = A_\text{GR}(f) + A_\text{EFT}$, where $A_\text{GR}$ is the amplitude of the IMRPhenomD model give in Eq.~\eqref{Eq. IMRPhD_Ins_amp}, while the EFT correction to the amplitude is 
\begin{align}
        A_\text{EFT}(f) & = \sqrt{\frac{5\pi}{24}}\frac{(GM)^2\sqrt{\eta}}{c^5 D_L}v^{-7/2} 
        \left[   -198\baw v_f^{10}
        \right. \nonumber \\
        &\left.  + 3\frac{22624\baq - (53149+16660\eta)\baw}{112} v_f^{12}  + {\cal{O}}(1/c^{13}) \right]\,.
\end{align}

Let us now focus on the Fourier phase in the SPA. Again, inserting Eqs.~(\ref{eq.E_tot})-(\ref{eq.E_a2}) and~(\ref{eq.Flux_tot})-(\ref{eq.Flux_a2}) into Eq.~(\ref{Eq. phase_SPA}) (or equivalently Eq.~\eqref{Eq. phase_SPA_2}), we find the EFT correction
\begin{align}
    \psi_{\text{EFT}} &= -\frac{351\baw}{8\eta}v^5 + \frac{1}{12544\eta} \left[  549360\baq 
    \right.\nonumber \\
    &\left. - 45(43683+12908\eta)\baw  \right] v^7 + {\cal{O}}(1/c^8)\,.
\end{align}

As previously stated, the flux in Eq.~\eqref{Eq. phase_SPA_2} and \eqref{eq.SPA_amp} are not exact to 1PN order, which means that our waveform is also not exact to this order. However, the missing terms play are expected to play a small role in our parameter estimation constraints on modified gravity, as shown in~\cite{Perkins:2022fhr}. Including these terms would probably improve our constraints by ${\cal{O}}(10\%)$, and thus, not including them leads to a conservative bound and will not change our conclusions.

\section{Extrapolation to estimate $a_\text{5PN}$ and $a_\text{6PN}$}
\label{sec:appendix_B}

We first use cubic spline interpolation to derive a relation between PN orders and PN coefficients. The cubic spline interpolation uses a series of cubic polynomials $S_i (x)$ to describe the relation $y(x)$ in the segment $[x_i, x_{i+1}]$, where $(x_i, y_i)$ and $(x_{i+1}, y_{i+1})$ are  neighboring data points. One then requires that $S_i(x_i) = S_{i-1} (x_i) = S_{i+1} (x_i) = y_i$ and that the function be $C(2)$-continuous to find the coefficients of the cubic polynomial.

\begin{figure}
    \centering
    \includegraphics[width=\columnwidth,clip=true]{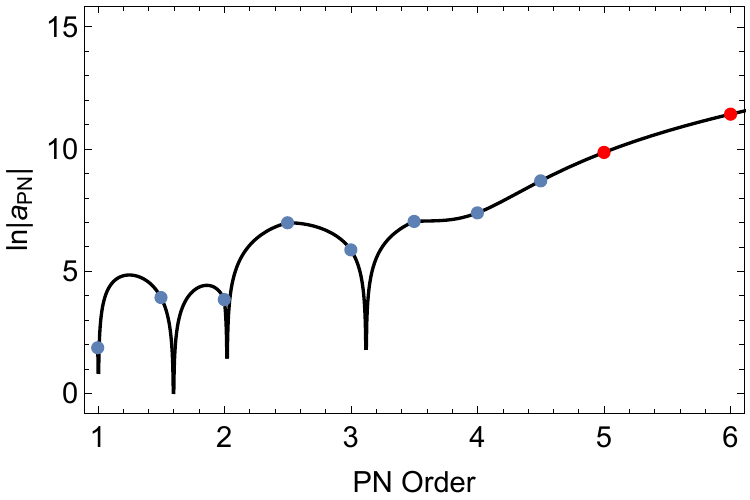}
    \caption{Interpolation function and result of extrapolation. The black line stands for the resulting function of cubic spline interpolation. The blue points are standard PN coefficients and red points are our extrapolated result for $a_\text{5PN}$ and $a_\text{6PN}$. 
    }
    \label{fig.PNfit}
\end{figure}

After interpolation, we use the last cubic polynomial to extrapolate and predict the 5PN and 6PN coefficients. The result is shown in Fig.~\ref{fig.PNfit}, where we present the cubic-spline interpolation (black), and the analytic results for the PN coefficients (blue dots). The cubic-spline evaluated at a PN order higher than 4 becomes an extrapolation, which we represent with red dots. These results are used in the main body of the test to set a prior on the magnitude of the EFT coefficients (see Eqs.~\eqref{Eq. PN_prior_2} and~\eqref{Eq. PN_prior_3}). 

\bibliography{ref}

\end{document}